\documentclass[lettersize,journal]{IEEEtran}
\usepackage{amsmath,amsfonts}
\usepackage{algorithmic}
\usepackage{algorithm}
\usepackage{array}
\usepackage[caption=false,font=normalsize,labelfont=sf,textfont=sf]{subfig}
\usepackage{textcomp}
\usepackage{stfloats}
\usepackage{url}
\usepackage{verbatim}
\usepackage{graphicx}
\usepackage{cite}
\usepackage{algorithmic}
\usepackage{array}
\usepackage{textcomp}
\usepackage{stfloats}
\usepackage{url}
\usepackage{verbatim}
\usepackage{graphicx}
\usepackage{algorithm}
\usepackage{algorithmic}
\usepackage{amssymb}
\usepackage{multirow}
\usepackage{booktabs}
\usepackage{amsmath,amsthm,amssymb,mathrsfs}
\usepackage{graphicx}
\usepackage[switch]{lineno}
\usepackage{xcolor}
\usepackage[nameinlink]{cleveref}
\usepackage{amsmath}

\theoremstyle{definition}
\newtheorem{definition}{Definition}[section]

\theoremstyle{plain}
\newtheorem{theorem}[definition]{Theorem}
\newtheorem{lemma}[definition]{Lemma}

\newcommand{\E}{\mathbb{E}}
\newcommand{\Var}{\mathrm{Var}}
\newcommand{\Cov}{\mathrm{Cov}}
\newcommand{\F}{\mathcal{F}}

\newcommand{\xlow}{x_{\mathrm{low}}}
\newcommand{\xhigh}{x_{\mathrm{high}}}
\newcommand{\xtilde}{\tilde{x}_0}
\newcommand{\xidr}{x_{\mathrm{IDR}}}
\begin{document}

\title{From Coarse to Continuous: Progressive Refinement Implicit Neural Representation for Motion-Robust Anisotropic MRI Reconstruction}

\author{Zhenxuan Zhang, Lipei Zhang, Yanqi Cheng, Zi Wang, Fanwen Wang, Haosen Zhang, Yue Yang, Yinzhe Wu, Jiahao Huang, Angelica I Aviles-Rivero, Zhifan Gao, Guang Yang, Peter J. Lally
% <-this % stops a space
\thanks{Guang Yang was supported in part by the ERC IMI (101005122), the H2020 (952172), the MRC (MC/PC/21013), the Royal Society (IEC/NSFC/211235), the NVIDIA Academic Hardware Grant Program, the SABER project supported by Boehringer Ingelheim Ltd, NIHR Imperial Biomedical Research Centre (RDA01), The Wellcome Leap Dynamic resilience program (co-funded by Temasek Trust)., UKRI guarantee funding for Horizon Europe MSCA Postdoctoral Fellowships (EP/Z002206/1), UKRI MRC Research Grant, TFS Research Grants (MR/U506710/1), Swiss National Science Foundation (Grant No. 220785), and the UKRI Future Leaders Fellowship (MR/V023799/1, UKRI2738). Imperial College London Seeds for Success Fund. Zhenxuan Zhang was supported by a CSC Scholarship. Guang Yang and Peter J. Lally are co-last authors. Corresponding author: Guang Yang (\texttt{g.yang@imperial.ac.uk}).}%
\thanks{Zhenxuan Zhang, Haosen Zhang, Yue Yang, Fanwen Wang, Yinzhe Wu, Jiahao Huang and Zi Wang are with the Department of Bioengineering and Imperial-X, Imperial College London, London SW7 2AZ, U.K.}%
\thanks{Peter J. Lally is with the Department of Bioengineering, Imperial College London, London SW7 2AZ, U.K. and the U.K. Dementia Research Institute Centre for Care Research \& Technology, London, W12 0BZ (\texttt{p.lally@imperial.ac.uk})}
\thanks{Lipei Zhang and Yanqi Cheng are with the Department of Applied Mathematics and Theoretical Physics, University of Cambridge, Cambridge CB3 0WA, U.K.}%
\thanks{Zhifan Gao is with the School of Biomedical Engineering, Sun
Yat-sen University, Shenzhen 518107, China}
\thanks{Angelica I. Aviles-Rivero is with the Yau Mathematical Sciences Center, Tsinghua University, Beijing 100084, China.}%
\thanks{Guang Yang is with the Bioengineering Department and Imperial-X, Imperial College London, London W12 7SL, U.K., and with the National Heart and Lung Institute, Imperial College London, London SW7 2AZ, U.K., and with the Cardiovascular Research Centre, Royal Brompton Hospital, London SW3 6NP, U.K., and with the School of Biomedical Engineering \& Imaging Sciences, King’s College London, London WC2R 2LS, U.K.}}
% <-this % stops a space

\markboth{Journal of \LaTeX\ Class Files,~Vol.~14, No.~8, August~2021}%
{Shell \MakeLowercase{\textit{et al.}}: A Sample Article Using IEEEtran.cls for IEEE Journals}


\maketitle

\begin{abstract}
In motion-robust magnetic resonance imaging (MRI), slice-to-volume reconstruction is critical for recovering anatomically consistent 3D brain volumes from 2D slices, especially under accelerated acquisitions or patient motion. However, this task remains challenging due to hierarchical structural disruptions. It includes local detail loss from k-space undersampling, global structural aliasing caused by motion, and volumetric anisotropy. Therefore, we propose a progressive refinement implicit neural representation (PR-INR) framework. Our PR-INR unifies motion correction, structural refinement, and volumetric synthesis within a geometry-aware coordinate space. Specifically, a motion-aware diffusion module is first employed to generate coarse volumetric reconstructions that suppress motion artifacts and preserve global anatomical structures. Then, we introduce an implicit detail restoration module that performs residual refinement by aligning spatial coordinates with visual features. It corrects local structures and enhances boundary precision. Further, a voxel continuous-aware representation module represents the image as a continuous function over 3D coordinates. It enables accurate inter-slice completion and high-frequency detail recovery. We evaluate PR-INR on five public MRI datasets under various motion conditions (3$\%$ and 5$\%$ displacement), undersampling rates (4x and 8x) and slice resolutions (scale = 5). Experimental results demonstrate that PR-INR outperforms state-of-the-art methods in both quantitative reconstruction metrics and visual quality. It further shows generalization and robustness across diverse unseen domains.
\end{abstract}

%{}
\section{Introduction}
\label{sec:introduction}
\begin{figure}[t]
\centerline{\includegraphics[width=\columnwidth]{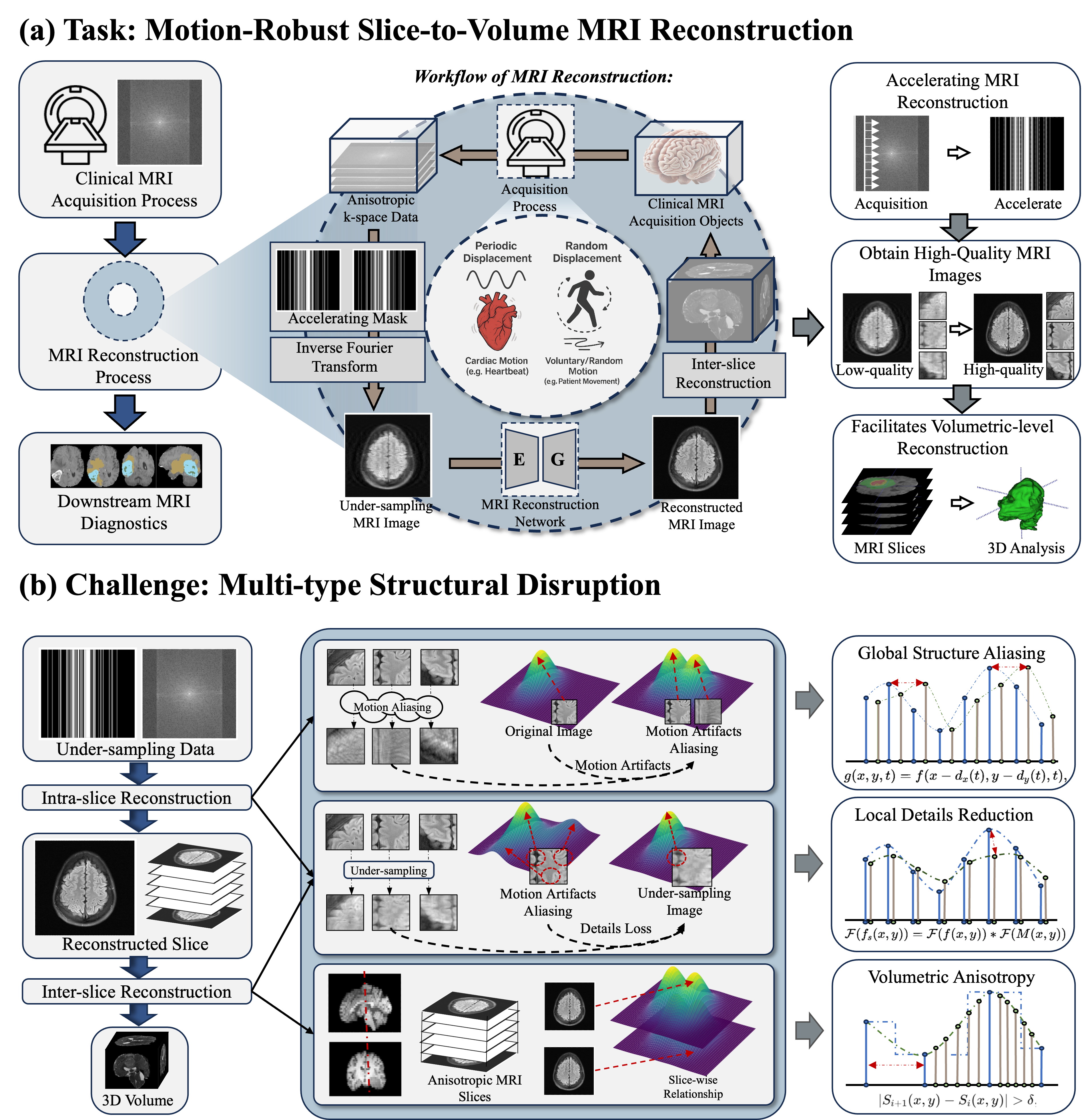}}
\caption{Overview of motion-robust slice-to-volume MRI reconstruction. (a) Illustrates the workflow from clinical MRI acquisition to inter-slice reconstruction and downstream volumetric diagnostics. (b) Highlights key challenge of multi-type structural disruptions, including global aliasing, local detail loss, and volumetric anisotropy.}
\label{fig1}
\end{figure}

\IEEEPARstart{M}{otion} robust anisotropic Magnetic Resonance Imaging (MRI) reconstruction is essential for accurate diagnosis and quantitative analysis across a wide range of clinical applications (Fig. \ref{fig1} (a)). On the one hand, many anatomically critical regions (e.g., the brain and heart) contain both fine-grained structural details and complex global organization. These require high-resolution imaging to visualize subtle pathological changes (e.g., microstructural damage in neurodegenerative diseases or small lesions in stroke) \cite{mri_review1}. On the other hand, it is also significant in reducing motion artifacts and aliasing \cite{moco_review,sobotka2022motion}.  Motion artifacts can obscure fine neural details and compromise the reliability of volumetric assessments, particularly in subjects like children and elderly patients \cite{sobotka2022motion,qiu2023medical}.
Further, robust slice-to-volume reconstruction is essential for synthesizing consistent 3D representations from 2D slices \cite{gholipour2010robust,rousseau2006registration,kim2009intersection,uus2020deformable}. It needs robust slice-to-volume reconstruction to ensure spatial consistency and enable downstream tasks \cite{gholipour2010robust,wang2024cross}. It allows precise volumetric analysis critical to understanding morphology and tracking disease progression \cite{wang2024cross}. Therefore, it not only enhances image quality but also facilitates accurate image biomarker extraction and quantification, which are important for downstream clinical tasks.

However, reconstructing volumetric MRI images from sparse motion-corrupted slices still has the challenge of multi-type structural disruptions (Fig. \ref{fig1}(b)). It can be divided into three aspects.
First, it still has the challenge of global structure aliasing. Motion during MRI acquisition introduces global artifacts (e.g., aliasing due to patient movement or respiration). These artifacts manifest as spurious signals that misrepresent anatomical boundaries or overlap regions \cite{kim2009intersection,pei2020anatomy,duffy2021retrospective}. This leads to mis-registration between structures when assessing large-scale connectivity or structural integrity \cite{alansary2017pvr,kainz2015fast}.
Second, it still has the challenge of local detail reduction. This is because of the loss of high spatial frequency information in accelerating MRI acquisition with incoherent undersampling schemes. It results in incomplete data and further leads to the loss of fine-grained local image details \cite{iglesias2021joint,liu20213d,lustig2007sparse,zhang2022soup}. This is particularly detrimental in capturing subtle anatomical structures (e.g., cortical folds or small subcortical regions). These structures are critical for accurate diagnosis and analysis \cite{wang2022adjacent,hou20183,salehi2018real}.
Third, it has the limitation of volumetric anisotropy. Clinical MRI acquisitions often have a high in-plane resolution but relatively low resolution along the slice dimension \cite{kuklisova2012reconstruction}. This disparity creates discontinuities and a lack of smoothness across slices, making the reconstruction of seamless 3D brain structures difficult (e.g., the gyri and sulci of the brain may appear disconnected or distorted). It limits the accuracy of neuroanatomical analyses or volumetric measurements.
These issues not only undermine the diagnostic quality but also hinder downstream applications. Therefore, it still requires a unified solution to address these challenges.

\begin{figure}[t]
\centerline{\includegraphics[width=\columnwidth]{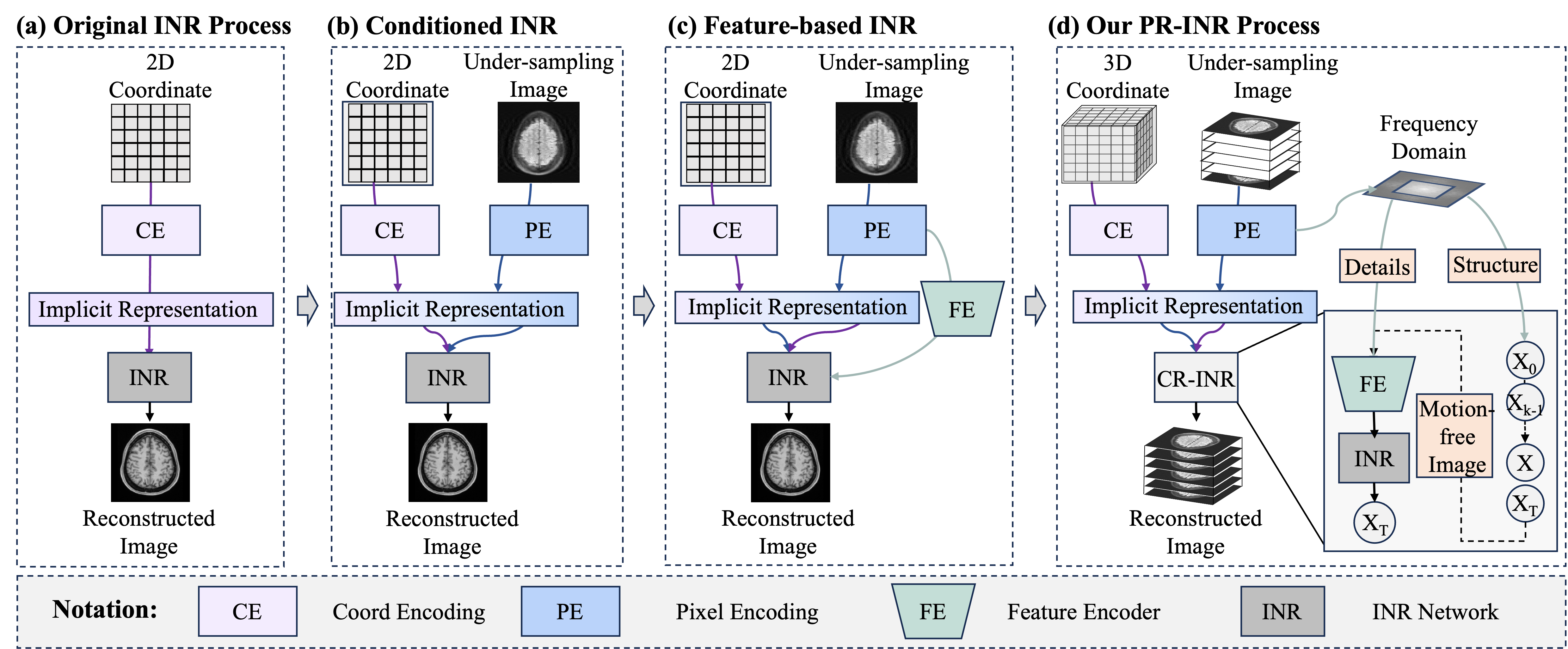}}
\caption{Overview of strategies integrating INR variants: (a) Original INR learns a continuous image solely from sinusoidally encoded spatial coordinates. (b) Conditioned INR injects a pixel prior to guide the coordinate stream. (c) Feature-based INR integrates a feature encoder. It iteratively refines the implicit representation together with extracted image features. (d) Proposed PR-INR generalizes to 3D coordinates and performs coarse-to-continuous reconstruction via a progressive architecture. It first removes motion artifacts and global aliasing with diffusion prior, then selectively restores high-frequency details through a residual INR, and finally enforces volumetric smoothness using a continuous 3D representation. This design yields motion-robust and anatomically consistent volumetric images.}
\label{fig2}
\end{figure}

Existing MRI reconstruction methods struggle to fully address these issues due to inherent limitations. Early approaches based on compressed sensing and conventional deep learning typically focus on sparsity priors \cite{lustig2007sparse,haldar2016p,mri_review1}. However, they often struggle under high acceleration factors due to the limited ability of sparsity priors to accurately complete missing information. CNN-based and generative networks can enhance perceptual quality under high acceleration factors but may introduce hallucinated details. It leads to anatomically implausible structures and compromised fidelity \cite{heckel2024deep}. Especially, these are often aggravated by motion in dynamic or long acquisition scenarios. Some recent methods integrate motion correction modules. But their performance heavily depends on the pre-defined motion priors \cite{tulyakov2018mocogan,wu2024anires2d}. Implicit neural representations (INR) have the potential to address these challenges for their continuous spatial structures and high-frequency sensitivity \cite{xu2022svort,young2024fully}. Early INR-based methods (Fig.~\ref{fig2}(a)) represent images as continuous functions over coordinates $\mathbf{x}_{\text{coord}}$, enabling the learning of implicit spatial priors. This allows for resolution-independent reconstruction and recovery of high-frequency details~\cite{xu2022svort,young2024fully}. However, they must be optimized per case and lack prior knowledge of measurements. Conditional INR (Fig. \ref{fig2}(b)) builds the network with  pixel priors $\mathbf{x_{pixel}}$. It allows fast inference but still suffers from inherited motion artifacts. Feature-based INR (Fig. \ref{fig2}(c)) further enhances robustness by extracting latent features $\mathbf{f}(\mathbf{x_{pixel}})$ from the input. But the fixed-single-stream INR still struggles with motion artifacts and anisotropic data. Therefore, it is still needed to combine the continuous modeling with the artifact suppression to enable robust recovery under motion and under-sampling.

In this work, we propose a unified progressive refinement implicit neural representation (PR-INR) framework for motion-robust slice-to-volume MRI reconstruction. The core idea of PR-INR is to treat MRI reconstruction as a task-disentangled process. It explicitly disentangles the reconstruction into progressive sub-tasks, including motion correction, detail recovery, and inter-slice continuity modeling. The key motivation is that structural misalignments caused by motion must be firstly addressed, as they interfere with subsequent detail restoration and volumetric consistency. Without a stabilized anatomical layout, local refinement may amplify noise or hallucinate misleading features. Therefore, PR-INR first performs motion-aware correction to produce a globally aligned coarse volume. This is followed by a residual implicit refinement stage that enhances high-frequency details, and a final stage that enforces slice-to-volume consistency in a continuous 3D coordinate space. Specifically, PR-INR first generates a coarse reconstruction result via a motion-aware diffusion (MAD) prior. It stabilizes motion artifacts and enforces global structure. This intermediate result is then passed through an implicit detail restoration (IDR) module that performs residual refinement. For each spatial query $(x,y)$ IDR concatenates a learned local feature vector, allowing the implicit field to be content–adaptive rather than purely positional.  The network predicts only a high-frequency residual that is weighted by a mask (i.e., enforcing data consistency). This utilizes the continuous function characteristics of INR to represent the multi-dimensional features of the image. Further, the voxel continuous-aware representation (VCR) module represents the reconstructed image as a continuous function over 3D coordinates and is conditioned on both spatial locations and local visual features. It enables accurate recovery of missing structures and high-frequency texture. In these processes, PR-INR achieves robust and high-fidelity reconstructions under varying slice thicknesses, motion artifacts and acceleration factors. In summary:
\begin{enumerate}
    \item We propose a unified reconstruction workflow that simultaneously addresses local detail loss, motion artifacts, and volumetric anisotropy.
    \item We design a novel reconstruction architecture that integrates motion-aware initialization, feature-conditioned implicit refinement, and continuous volumetric representation for unified MRI reconstruction.
    \item We conduct extensive experiments on five public MRI datasets, demonstrating superior performance across various sampling patterns and motion conditions.
\end{enumerate}

\section{Related Work}

\subsection{MRI Reconstruction.}
MRI reconstruction has traditionally relied on model-based techniques such as compressed sensing and parallel imaging \cite{lustig2007sparse,haldar2016p}. These methods impose handcrafted priors like sparsity or coil sensitivity, but often degrade at high acceleration rates. \textcolor[RGB]{0,0,0}{Recently, unsupervised and self-supervised reconstruction paradigms have gained increasing attention. These methods formulate reconstruction through measurement consistency or cooperative constraints, enabling effective recovery from incomplete data without paired supervision \cite{feng2025spatiotemporal,quan2024siamese}. However, these methods rely on specific redundancy assumptions and do not explicitly address motion-induced artifacts.} Motion artifacts further compromise image quality \cite{moco_review,kim2009intersection}. While prospective gating or navigator-based regression can reduce motion during acquisition, they are not always feasible (e.g., increasing acquisition time or requiring additional hardware)\cite{Pilot_Tone}. Retrospective correction strategies (e.g., learning-based or motion-compensated reconstruction) have been explored to restore spatial fidelity after motion has occurred \cite{qiu2023medical,lei2024joint}. These methods enhance structural clarity by recovering sharp features obscured by motion. More recently, INR models have been applied to volumetric reconstruction from 2D slices \cite{videoinr,sainr,cycleinr}. These approaches can improve inter-slice consistency by modeling continuous spatial fields.

\subsection{Diffusion-based methods.}
Diffusion-based methods model the MRI reconstruction task as a stochastic denoising process. It progressively transforms noisy observations toward fully sampled images. In MRI reconstruction, diffusion priors can be imposed either in the image domain or directly in the frequency domain, guiding reconstructions through learned score functions that capture the underlying data distribution \cite{ li2024rethinking}. Recent approaches utilize conditional diffusion models to incorporate measurement consistency at each diffusion step, blending the forward acquisition operator with denoising updates to ensure fidelity to the raw $k$-space data \cite{gungor2023adaptive,chung2022score}. Latent diffusion frameworks further improve efficiency by mapping high-dimensional image volumes into a lower-dimensional latent space. This diffusion process operates more tractably before decoding back to the full-resolution image \cite{ozturkler2023smrd,li2024rethinking}. These strategies have been shown to enhance robustness to noise and motion, yielding higher PSNR and sharper anatomical details \cite{li2024rethinking}. However, diffusion models carry the risk of generating hallucinations (e.g., synthesizing anatomically plausible but incorrect structures). It can compromise clinical reliability \cite{mri_review2, li2024rethinking}. 

\subsection{Implicit Neural Representation.}
In contrast to discretized grid-based methods, implicit neural representations (INRs) treat an image or volume as a continuous function mapping spatial coordinates to intensities. Early INR methods struggled with high-frequency details, leading to advanced designs incorporating sinusoidal activations (e.g., SIREN \cite{siren}) or Fourier feature embeddings \cite{videoinr}. In MRI, INRs have been utilized for super-resolution and slice-to-volume reconstruction \cite{INR_MRI1,INR_MRI2,INR_MRI3}, promising superior resolution and inter-slice continuity without large memory overhead. More recent efforts embed spatial attention or motion modeling into INR frameworks \cite{sainr}, making them robust to undersampled and corrupted $k$-space data. Although INR-driven reconstructions can excel at preserving details, they risk hallucinating structures if not tightly coupled with measurement fidelity, especially under motion-induced misalignment.

\begin{figure*}[t]
\centerline{\includegraphics[width=2.0\columnwidth]{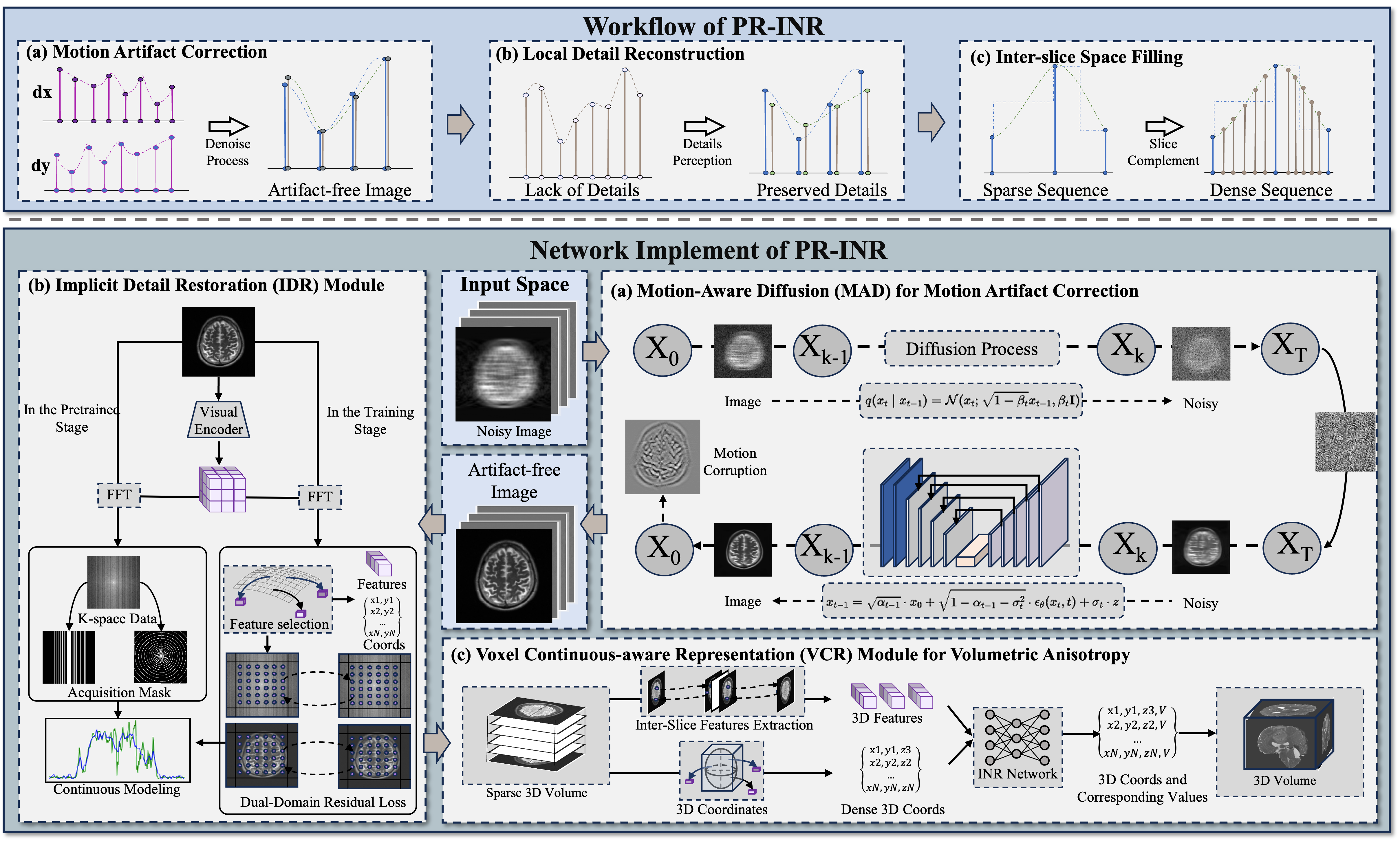}}
\caption{Overview of the proposed PR-INR framework for motion-robust slice-to-volume MRI reconstruction.  (a) Motion artifact correction. A motion-aware diffusion-based model is used to suppress intra-slice artifacts by iteratively refining image quality through forward and reverse denoising processes. (b) Local detail reconstruction. A data-consistency-aware visual encoder extracts anatomical details from under-sampled slices, guided by k-space priors and acquisition-aware feature selection. The residual INR then selectively refines high-frequency details to correct hallucinations introduced during the diffusion phase. It ensures spatial fidelity and structural consistency. (c) Inter-slice space filling. A volumetric INR takes continuous spatial coordinates and encoded features to generate dense volumetric outputs with coherent inter-slice structure.}
\label{fig3}
\end{figure*}

\begin{table}[t]
\centering
\caption{Glossary of Symbols}
\label{tab:glossary}
\resizebox{\columnwidth}{!}{
\begin{tabular}{ll|ll}
\toprule
\textbf{Symbol} & \textbf{Description} & \textbf{Symbol} & \textbf{Description} \\
\midrule
$\mathbf{k}$ & Fully-sampled $k$-space & 
$\mathbf{b}$ & Undersampled $k$-space \\
$\hat{\mathbf{k}},\ \hat{\mathbf{x}}$ & Reconstructed $k$-space / image & 
$\mathbf{x}$ & Ground-truth image \\
$\widetilde{\mathbf{x}}$ & Motion-corrupted image & 
$\tilde{x}_0$ & Data-consistent corrected image \\
$\mathcal{F}$ & Fourier transform & 
$\mathbf{P}$ & $k$-space sampling mask \\
$x_{\text{MAD}}$ & Motion-aware diffusion output & 
$x_{\text{IDR}}$ & Output of detail restoration module \\
$\boldsymbol{\eta}$ & Measurement noise & 
$\Omega$ & Imaging spatial domain \\
$W$ & Frequency weighting mask & 
$M,\ N$ & Number of slices / pixels \\
$\tau_j$ & Motion transformation at line $j$ & 
$\delta^*$ & Estimated motion field \\
$z$ & Standard Gaussian noise & 
$\mathcal{R}(\cdot)$ & Regularization prior \\
$x_t$ & Noisy image at step $t$ & 
$\bar{\alpha}_t,\ \sigma_t$ & Noise schedule coefficients \\
$f_{\theta_1}$ & Motion correction network & 
$\epsilon_\theta$ & Denoiser in diffusion model \\
$f_{\theta_2}$ & INR for detail refinement & 
$r(x,y)$ & High-frequency residual \\
$f_{\theta_3}$ & Volumetric INR over $(x,y,z)$ & 
$\gamma(\mathbf{x})$ & Frequency-encoded coordinates \\
$\mathbf{c_1}$ & Conditioning visual features (2D) & 
$\mathbf{c_2}$ & Conditioning visual features (3D) \\
\bottomrule
\end{tabular}}
\end{table}

\section{Methodology}
\subsection{Problem Formulation}

Let $\mathbf{b} \in \mathbb{C}^M$ denote the observed undersampled $k$-space measurements, and $\mathbf{k} \in \mathbb{C}^N$ represent the fully-sampled $k$-space, where $M \ll N$. The corresponding fully-sampled image is given by $\mathbf{x} = \mathcal{F}^{-1} \mathbf{k}$, where $\mathcal{F}^{-1}$ denotes the inverse Fourier transform. The acquisition model is expressed as:
\begin{equation}
\mathbf{b} = \mathbf{P} \mathbf{k} + \boldsymbol{\eta}, 
\end{equation}
where $\mathbf{P}$ is the undersampling mask and $\boldsymbol{\eta} \in \mathbb{C}^M$ models measurement noise. The goal is to recover the fully-sampled $k$-space $\hat{\mathbf{k}}$, and reconstruct the corresponding image $\hat{\mathbf{x}} = \mathcal{F}^{-1} \hat{\mathbf{k}}$. This problem is typically formulated as a regularized inverse problem:
\begin{equation}
\min_{\hat{\mathbf{k}}} \frac{1}{2}\| \mathbf{P} \hat{\mathbf{k}} - \mathbf{b} \|_2^2 + \lambda \mathcal{R}(\hat{\mathbf{k}}), 
\end{equation}
where the first term ensures data consistency, and $\mathcal{R}(\hat{\mathbf{k}})$ serves as a prior to compensate for missing information due to undersampling.

\subsubsection{Global Structure Aliasing due to Motion}
In acquisition, motion-induced aliasing arises when different regions of k-space (e.g. phase encoding lines) are sampled under varying spatial transformations. \textcolor[RGB]{0,0,0}{In this work, we focus on slice-based 2D Cartesian acquisitions with thick slices. Under this anisotropic acquisition regime, spatial encoding is dominated by in-plane resolution, while the through-plane resolution is substantially coarser due to the large slice thickness. Under this setting, the effect of motion-corrupted sampling can be approximated as:}
\begin{equation}
\widetilde{\mathbf{x}} = \mathcal{F}^{-1}[\textstyle{\sum_{j}}[\mathcal{F}_j\left( \mathbf{x} \circ \tau_j \right)]] + \boldsymbol{\eta}
\label{equ3}
\end{equation}
where $\mathcal{F}_j$ denotes the $j$-th phase-encoding line of the 2D Fourier transform,  \textcolor[RGB]{0,0,0}{$\tau_j$ represents the in-plane rigid transformation associated with subject motion during the acquisition of the $j$-th line}, 
and $\boldsymbol{\eta}$ denotes measurement noise.
\textcolor[RGB]{0,0,0}{This formulation captures the dominant impact of in-plane motion on slice-based 2D acquisitions.}

\textcolor[RGB]{0,0,0}{It note that Eq.~(\ref{equ3}) does not aim to explicitly model kz-dependent through-plane encoding, slice profile effects, or fully acquisition-consistent 3D motion processes. Instead, it serves as a practically relevant abstraction that isolates the dominant motion-induced aliasing effects under thick-slice acquisition, while maintaining computational tractability. The limitations of this abstraction are discussed in Discussion Section.}

To compensate for such corruption, motion artifact correction can be formulated as a motion-aware image regression problem. Specifically, one seeks to predict the motion-corrected image intensity at aligned spatial coordinates by minimizing:
\begin{equation}
\theta_1^* = \arg\min_{\theta_1} \frac{1}{N} \sum_{i=1}^N \left| f_{\theta_1}(\widetilde{\mathbf{x}}) - \widetilde{y}_i \right|^2,
\end{equation}
\textcolor[RGB]{0,0,0}{where $\widetilde{y}_i$ denotes the motion-free reference image intensity under the same slice-based acquisition setting.} This formulation characterizes motion artifact correction as the recovery of spatially consistent image content from motion-corrupted slice-based observations.

\subsubsection{High-frequency Detail Loss due to Undersampling}

Even after motion correction, incoherent undersampling in the frequency domain introduces additional artifacts, primarily manifesting as incoherent aliasing and high-frequency detail loss. The degraded observation can be expressed as:
\begin{equation}
\hat{\mathbf{x}} = \mathcal{F}^{-1} \left[ \mathbf{P} \cdot \textstyle{\sum_{j}}[\mathcal{F}_j\left( \mathbf{x} \circ \tau_j \right)] \right] + \boldsymbol{\eta},
\end{equation}
where $\mathbf{P}$ denotes the undersampling mask. The task of restoring high-frequency anatomical structures can be formulated as a detail enhancement problem:
\begin{equation}
\theta_2^* = \arg\min_{\theta_2} \frac{1}{N} \sum_{i=1}^N \left| f_{\theta_2} \left( \hat{\mathbf{x}} \right) - \hat{y_i} \right|^2,
\end{equation}
where $\hat{y_i}$ represents the fully-sampled motion-free image intensity at location $\mathbf{x}_i$. This step aims to recover the fine-grained structures that are degraded due to frequency domain undersampling.

\subsubsection{Volumetric Anisotropy}
Clinical MRI data typically suffer from resolution anisotropy, where the through-plane resolution ($\Delta z$) is significantly lower than the in-plane resolutions ($\Delta x$, $\Delta y$). This results in inconsistent inter-slice continuity and volumetric degradation. The anisotropic acquisition can be characterized by the imaging domain: $(x,y,z) \in \Omega, \quad \Omega = \left[0, N_x \Delta x\right] \times \left[0, N_y \Delta y\right] \times \left[0, N_z \Delta z\right].$ To address this, volumetric consistency is commonly formulated as a 3D interpolation task using a continuous function defined over the anisotropic coordinate system:
\begin{equation}
\theta_3^* = \arg\min_{\theta_3} \frac{1}{M} \sum_{i=1}^{M} \int_{\Omega_{xy}} \left| f_{\theta_3}(x,y,z_i) - f_{\theta_2}(x,y,z_i) \right|^2 dx\,dy,
\end{equation}
where $\Omega_{xy}$ denotes the in-plane pixel intensities, $z_i$ indexes the available slices and M is the total number of observed slices. This formulation reflects the need to interpolate across sparsely sampled slices while preserving in-plane details.

\subsection{Network Implementation}

\begin{algorithm}[t]
\caption{Progressive Refinement Training of PR-INR}
\begin{flushleft}
\textbf{Input:} Undersampled $k$-space $y$, sampling mask $M$, Fourier operator $\mathcal{F}$, diffusion denoiser $\epsilon_{\theta_1}$, detail INR $f_{\theta_2}$, volumetric INR $f_{\theta_3}$, encoder $c(\cdot)$ \\
\textbf{Output:} Final volumetric reconstruction $x_{\text{VCR}}$
\end{flushleft}
\begin{algorithmic}[1]
\FOR{each training iteration}
    \STATE \textbf{Stage 1: Motion-aware Diffusion}
    \STATE Sample $x_T \sim \mathcal{N}(0,I)$
    \FOR{$t = T, T{-}1, \ldots, 1$}
        \STATE $\hat{\epsilon}_{\theta_1} \leftarrow \epsilon_{\theta_1}(x_t, t)$
        \STATE $x_{\text{MAD}} \leftarrow (x_t - \sqrt{1 - \bar{\alpha}_t} \hat{\epsilon}_{\theta_1})/{\sqrt{\bar{\alpha}_t}}$
        \STATE $x_{t-1} \leftarrow \sqrt{\bar{\alpha}_{t-1}} x_{\text{MAD}} + \sqrt{1 - \bar{\alpha}_{t-1} - \sigma_t^2} \hat{\epsilon}_{\theta_1} + \sigma_t z$
    \ENDFOR

    \STATE \textbf{Stage 2: Implicit Detail Restoration}
    \STATE $\tilde{x}_0 \gets \mathcal{F}^{-1} \left( M \odot y + (1 - M) \odot \mathcal{F}(x_{\text{MAD}}) \right)$
    \STATE // Ensures $\tilde{x}_0$ agrees with measured $k$-space at sampled locations
    \FOR{each pixel $(x, y)$}
        \STATE $r(x, y) \gets f_{\theta_2}(x, y;\, c_1(\tilde{x}))$
        \STATE $x_{\text{IDR}}(x, y) \gets \tilde{x}_0(x, y) + r(x, y)$
    \ENDFOR

    \STATE \textbf{Stage 3: Voxel Continuous-aware Volumetric}
    \STATE $\mathbf{c}_2 \leftarrow c(x_{\text{IDR}})$ \hfill // Aggregate 3D implicit features
    \STATE $x_{\text{VCR}}(x,y,z) \leftarrow f_{\theta_3}(x,y,z;\mathbf{c}_2(x,y,z))$
    \STATE $\mathcal{L}_{total} \leftarrow \mathcal{L}_{\text{MAD}} + \mathcal{L}_{\text{IDR}} + \mathcal{L}_{\text{VCR}}$\hfill // Loss Aggregation
    \STATE Update all parameters $\theta_1,\theta_2,\theta_3$ via backpropagation
\ENDFOR
\end{algorithmic}
\end{algorithm}

\subsubsection{Motion-Aware Diffusion Module for Denoising Motion Corruption}

Although the estimated motion field $\delta^*$ enables partial correction of motion-induced aliasing as formulated in Equation. (\ref{equ3}), residual errors often persist due to limited sampling density and imperfect motion alignment. These residuals degrade high-frequency details of the motion-compensated image.

To address this, a Motion-Aware Diffusion Module (MAD) is considered, where the partially corrected image is treated as a noisy observation (Fig. \ref{fig3}(a)). {The MAD module follows a diffusion-based U-Net backbone with a depth of 4, using [64, 128, 256, 512] filters per stage and SiLU activations. It takes 4-channel inputs with 32 base channels and employs channel multipliers of [1, 2, 4, 8] to capture multi-scale motion features. The conditional embedding dimension is set to 128. It progressively refining the diffusion prior under motion-aware conditioning. }The forward process progressively adds Gaussian noise to the motion-corrected image $x_0$:
\begin{equation}
q(x_t|x_0) = \mathcal{N}\left( x_t; \sqrt{\bar{\alpha}_t} x_0, (1 - \bar{\alpha}_t)\mathbf{I} \right), 
\end{equation}
where $x_0$ is reconstructed from the motion-aligned $k$-space obtained via Equation. (\ref{equ3}), $\{\beta_s\}_{s=1}^T$ is a pre-defined variance schedule, $\alpha_s=1-\beta_s$, and $\bar{\alpha}_t=\prod_{s=1}^t \alpha_s$.

The diffusion model is tasked to restore motion-free structures by minimizing the following denoising objective:
\begin{equation}
\mathcal{L}_{\text{MAD}} = \mathbb{E}_{x_0, t, \epsilon} \left[ \left\| \epsilon - \epsilon_\theta\left(x_t, t, \delta^*\right) \right\|^2 \right], 
\end{equation}
where $\epsilon_\theta$ predicts the noise residual conditioned on the corrupted image $x_t$, the timestep $t$, and the estimated motion field $\delta^*$. At inference, the clean motion-free image is iteratively reconstructed via the reverse diffusion process:
\begin{equation}
x_{t-1} = \sqrt{\bar{\alpha}_{t-1}}\, {x}_{\text{MAD}} + \sqrt{1 - \bar{\alpha}_{t-1} - \sigma_t^2} \hat{\epsilon}_\theta + \sigma_t z, 
\end{equation}
with $z\sim\mathcal{N}(0,I)$, and
\begin{equation}
{x}_{\text{MAD}} = (x_t - \sqrt{1 - \bar{\alpha}_t} \hat{\epsilon}_\theta)/({\sqrt{\bar{\alpha}_t}}). 
\end{equation}

\begin{figure*}[t]
\centerline{\includegraphics[width=2.05\columnwidth]{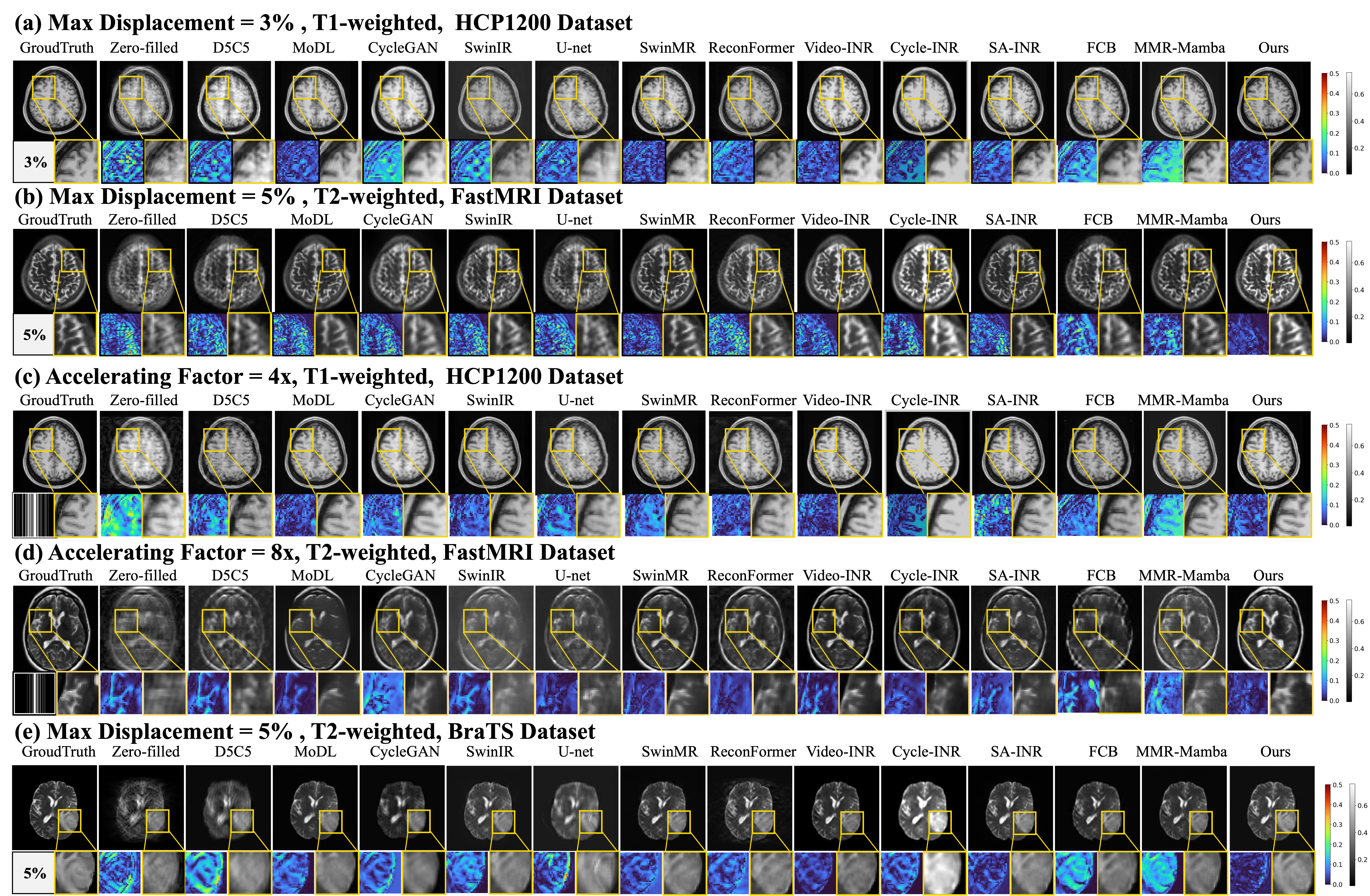}}
\caption{Visual comparisons of MRI reconstruction methods under different acceleration factors (AF) and motion displacements. Each subfigure shows reconstructed results from different methods. (a) Reconstruction under Max Displacement = 3 $\%$. (b) Reconstruction under Max Displacement = 5 $\%$. (c) Reconstruction with AF = 4x. (d) Reconstruction with AF = 8x. (e) Reconstruction under Max Displacement = 5 $\%$, T2-weighted, BraTS dataset. Each row includes reconstructed images (top) and error maps (bottom). }
\label{fig4}
\end{figure*}
\begin{figure}[t]
\centerline{\includegraphics[width=\columnwidth]{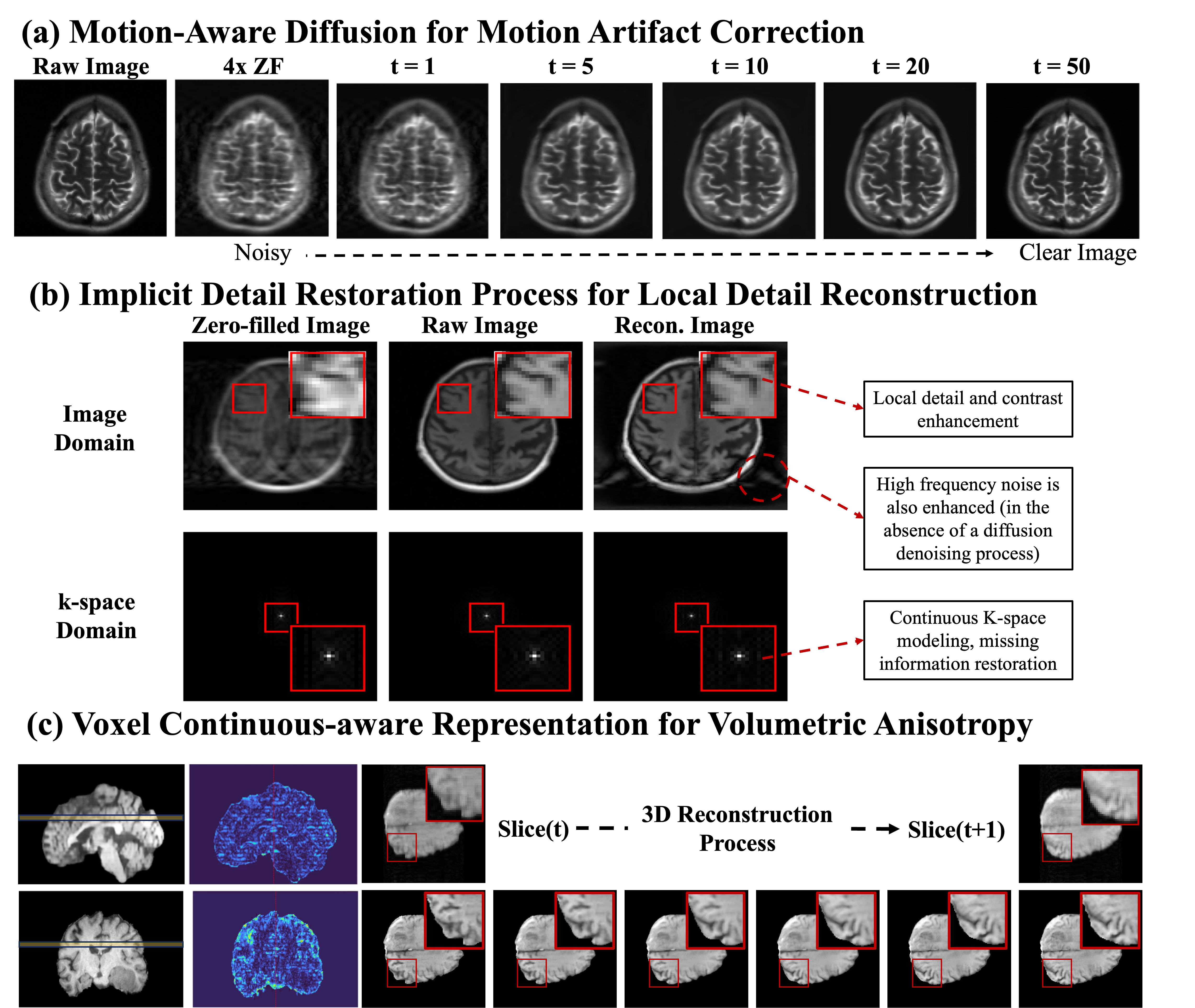}}
\caption{Overview of the proposed multi-stage MRI reconstruction framework.
(a) The motion aware diffusion process progressively removes motion artifacts and restores clean anatomical structures. (b) The implicit detail restoration process enhances local structural details and contrast in both image and k-space domains. (c) The voxel continuous-aware representation captures slice-wise continuity and reconstructs anisotropic volumes.}
\label{fig5}
\end{figure}
 \begin{figure}[t]
\centerline{\includegraphics[width=\columnwidth]{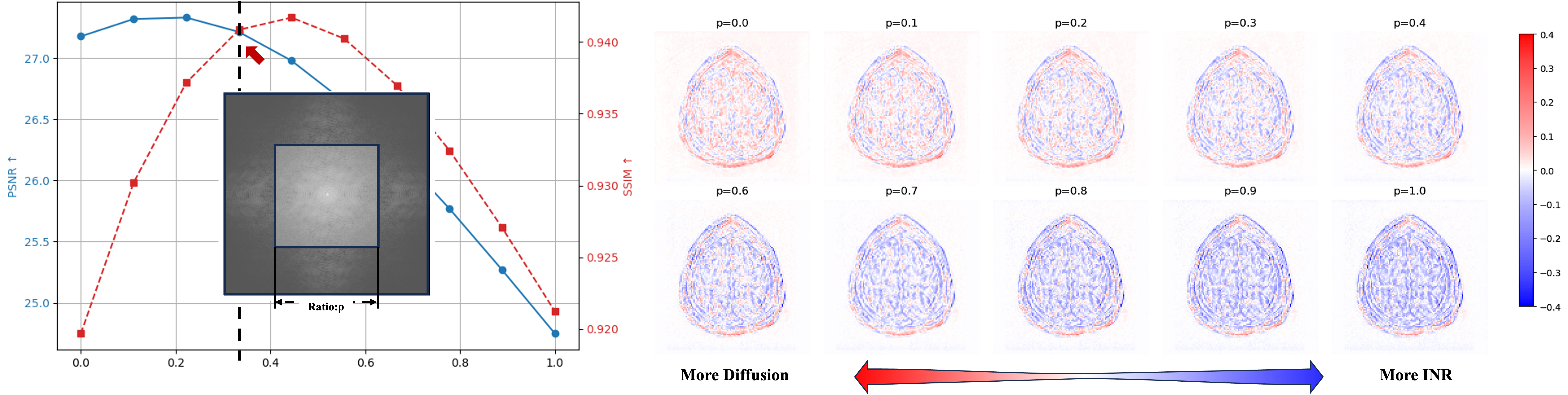}}
\caption{Visualization of the hybrid reconstruction strategy in the frequency domain. The left panel shows the balance of the INR model (high frequency) and the diffusion model (low frequency) determined by the $\rho$ in the frequency domain. This shows how the PSNR and SSIM change with different $\rho$. The right panel displays error maps for varying $\rho$, transitioning from full diffusion ($\rho=0.0$) to full INR ($\rho=1.0$).}
\label{fig6}
\end{figure}

This formulation characterizes the residual motion corruption removal problem as a conditional generative denoising task, where $\delta^*$ serves as the guidance derived from the estimated motion field.

\subsubsection{Implicit Detail Restoration Module for High-Frequency Enhancement}

Although diffusion models provide plausible completions in under-sampled regions, they are prone to \emph{hallucinations}, especially for high-frequency details that are weakly constrained by the observed frequency domain. To address this, we incorporate an Implicit Detail Restoration Module (IDR) conditioned on diffusion outputs and measured data (Fig. \ref{fig3}(b)). {The IDR module is implemented as a conditional implicit neural representation that models slice-wise fine details from spatial coordinates and diffusion features. It contains four hidden layers with 256 neurons each, using sine activations and frequency scaling parameters $\omega_0^{(1)} = 30$ and $\omega_0^{(h)} = 30$ to encode high-frequency structures effectively. The module takes as input the 2D spatial coordinates along with contextual embeddings derived from diffusion priors, and predicts residual corrections in both the image and frequency domains. }

We first apply a data-consistency projection to correct diffusion's low-frequency bias:
\begin{equation}
\tilde{x}_0 = \mathcal{F}^{-1}\left[ P \odot y + (1 - P) \odot \mathcal{F}(x_{\text{MAD}}) \right].
\end{equation}
This projection enforces:
\begin{equation}
\mathcal{F}(\tilde{x}_0)\big|_{P=1} = y\big|_{P=1},
\end{equation}
thus guaranteeing that the low-frequency spectrum strictly conforms to measurements. However, $\tilde{x}_0$ still lacks reliable high-frequency content due to undersampling. 
{Here, the data-consistency projection replaces the acquired k-space coefficients with ground-truth measurements while preserving the diffusion-estimated values in unmeasured regions, ensuring physical fidelity to the acquisition process. Although this correction primarily targets the low-frequency bias introduced by the diffusion prior, the inverse Fourier transform inherently couples frequency components in the spatial domain. As a result, restoring accurate low-frequency coefficients improves global contrast consistency, which stabilizes the reconstruction of adjacent high-frequency regions and reduces spurious edge artifacts. This coupled correction effect provides a smooth spectral transition that benefits the subsequent high-frequency refinement in the IDR module.
}

To selectively refine only the high-frequency residual, we adopt a residual INR model:
\begin{equation}
{x}_{\text{IDR}}(x,y) = \tilde{x}_0(x,y) + f_{\theta_2}\left( x,y; \mathbf{c}_1(x,y) \right),
\end{equation}
where $f_{\theta_2}$ is a conditioned INR responsible for learning, the conditioning vector \(\mathbf{c}_1(x,y)\) is extracted via a 2D convolutional encoder applied to \(\tilde{x}_0\):
\begin{equation}
r(x,y) = {x}_{\text{IDR}}(x,y) - \tilde{x}_0(x,y).
\end{equation}
This residual formulation $r(x,y)$ effectively restricts $f_{\theta_2}$ to focus on hallucination-prone, high-frequency components.

Unlike convolutional layers that favor low-frequency signals, SIREN-based INRs use periodic activations:
\begin{equation}
f_{\theta_2}(\mathbf{x}) = \sin(\omega_0 \cdot \mathbf{W} \cdot \gamma(\mathbf{x}) + \mathbf{b}),
\end{equation}
where \(\mathbf{W}\) and \(\mathbf{b}\) are the learnable weight and bias parameters of the INR layer, and \(\omega_0\) is a hyperparameter that controls the frequency scale of the sinusoidal activation. $\gamma(\mathbf{x}) = [\sin(\mathbf{B}\mathbf{x}),\ \cos(\mathbf{B}\mathbf{x})]$ as the frequency encoding, where $\mathbf{B}$ is a frequency mapping matrix that controls the spectral bandwidth of the encoding. This formulation allows the INR to represent fine-grained, high-frequency anatomical structures.

\begin{table*}[t]
\centering
\caption{Comparison with the state-of-the-art MRI reconstruction methods on fastMRI and HCP1200 datasets under motion corruption. 'DM' denotes the displacement magnitude in pixels. The 'AF' denotes the accelerating factor. The values in brackets '(*)' represent the standard deviation. Best results are in bold.}
{
\label{tab:motion}
\resizebox{\textwidth}{!}{
\begin{tabular}{ll|l|rrrr|rrrr}
\hline
\multicolumn{3}{c|}{} & \multicolumn{4}{c|}{fastMRI} & \multicolumn{4}{c}{HCP1200} \\ 
\hline
\multicolumn{1}{c|}{DM} & \multicolumn{1}{c|}{Methods} & \multicolumn{1}{c|}{Years}
& PSNR$\uparrow$ & SSIM$\uparrow$ & LPIPS$\downarrow$ & NCC$\uparrow$
& PSNR$\uparrow$ & SSIM$\uparrow$ & LPIPS$\downarrow$ & NCC$\uparrow$ \\
\hline\hline
\multicolumn{1}{c|}{} & Zero-filled & - & 24.3 (4.8) & 0.599 (0.126) & 0.204 (0.050) & 0.917 (0.095) & 20.5 (3.5) & 0.674 (0.109) & 0.242 (0.064) & 0.934 (0.023) \\
\multicolumn{1}{c|}{} & D5C5 \cite{D5C5} & 2017 & 27.0 (4.7) & 0.697 (0.121) & 0.267 (0.078) & 0.950 (0.088) & 21.6 (3.7) & 0.804 (0.063) & 0.181 (0.045) & 0.968 (0.011) \\
\multicolumn{1}{c|}{} & MoDL \cite{MoDL} & 2018 & 32.4 (3.8) & 0.829 (0.120) & 0.135 (0.075) & 0.972 (0.107) & 22.7 (2.8) & 0.702 (0.181) & 0.194 (0.026) & 0.974 (0.021) \\
\multicolumn{1}{c|}{} & CycleGAN \cite{Cyclegan} & 2017 & 26.2 (3.5) & 0.746 (0.130) & 0.268 (0.075) & 0.958 (0.088) & 19.5 (4.2) & 0.727 (0.121) & 0.137 (0.036) & 0.969 (0.048) \\
\multicolumn{1}{c|}{} & SwinIR \cite{swinir} & 2021 & 29.0 (4.5) & 0.793 (0.095) & 0.191 (0.042) & 0.971 (0.053) & 23.2 (4.3) & 0.802 (0.064) & 0.135 (0.037) & 0.981 (0.008) \\
\multicolumn{1}{c|}{} & U-net \cite{unet} & 2015 & 25.7 (3.5) & 0.678 (0.193) & 0.274 (0.094) & 0.960 (0.073) & 20.9 (2.9) & 0.670 (0.179) & 0.214 (0.091) & 0.953 (0.047) \\
\multicolumn{1}{c|}{} & SwinMR \cite{swinmr} & 2022 & 32.7 (4.4) & 0.871 (0.066) & 0.105 (0.028) & 0.984 (0.037) & 26.6 (3.8) & 0.838 (0.057) & 0.098 (0.034) & 0.986 (0.009) \\
\multicolumn{1}{c|}{3$\%$} & ReconFormer \cite{reconformer} & 2024 & 25.6 (4.1) & 0.617 (0.089) & 0.291 (0.036) & 0.941 (0.073) & 25.1 (4.0) & 0.820 (0.069) & 0.116 (0.025) & 0.986 (0.006) \\
\multicolumn{1}{c|}{} & Video-INR \cite{videoinr} & 2022 & 32.8 (3.9) & 0.837 (0.078) & 0.141 (0.046) & 0.974 (0.062) & 26.4 (4.0) & 0.928 (0.049) & 0.069 (0.018) & 0.984 (0.004) \\
\multicolumn{1}{c|}{} & CycleINR \cite{cycleinr} & 2024 & 26.0 (4.0) & 0.805 (0.110) & 0.206 (0.068) & 0.969 (0.063) & 20.4 (3.5) & 0.827 (0.091) & 0.149 (0.045) & 0.964 (0.049) \\
\multicolumn{1}{c|}{} & SA-INR \cite{sainr} & 2024 & \textbf{33.9 (4.2)} & 0.886 (0.065) & 0.083 (0.032) & 0.987 (0.034) & 26.7 (3.4) & 0.925 (0.056) & 0.061 (0.024) & 0.990 (0.006) \\
\multicolumn{1}{c|}{} & FCB \cite{FCB} & 2025 & 27.2 (4.3) & 0.830 (0.092) & 0.170 (0.070) & 0.962 (0.056) & 23.4 (4.5) & 0.872 (0.069) & 0.116 (0.037) & 0.983 (0.007) \\
\multicolumn{1}{c|}{} & MMR-Mamba \cite{MMR-Mamba} & 2025 & 26.6 (5.9) & 0.846 (0.115) & 0.124 (0.081) & 0.975 (0.058) & 23.9 (6.0) & 0.851 (0.053) & 0.069 (0.034) & 0.989 (0.009) \\
\multicolumn{1}{c|}{} & Our PR-INR & 2025 & 33.3 (2.3) & \textbf{0.919 (0.035)} & \textbf{0.063 (0.029)} & \textbf{0.990 (0.007)} & \textbf{27.0 (3.3)} & \textbf{0.934 (0.064)} & \textbf{0.060 (0.009)} & \textbf{0.995 (0.009)} \\

\hline
\multicolumn{1}{c|}{} & Zero-filled & - & 21.7 (4.6) & 0.456 (0.144) & 0.278 (0.068) & 0.864 (0.123) & 18.3 (2.9) & 0.510 (0.146) & 0.342 (0.077) & 0.895 (0.032) \\
\multicolumn{1}{c|}{} & D5C5 \cite{D5C5} & 2017 & 24.9 (4.8) & 0.614 (0.129) & 0.315 (0.065) & 0.932 (0.097) & 18.3 (3.7) & 0.708 (0.089) & 0.261 (0.055) & 0.943 (0.016) \\
\multicolumn{1}{c|}{} & MoDL \cite{MoDL} & 2018 & 30.6 (3.6) & 0.778 (0.148) & 0.172 (0.094) & 0.963 (0.128) & 19.6 (3.9) & 0.613 (0.153) & 0.266 (0.032) & 0.945 (0.085) \\
\multicolumn{1}{c|}{} & CycleGAN \cite{Cyclegan} & 2017 & 25.3 (3.4) & 0.714 (0.138) & 0.281 (0.074) & 0.949 (0.100) & 17.3 (3.8) & 0.634 (0.124) & 0.189 (0.046) & 0.954 (0.052) \\
\multicolumn{1}{c|}{} & SwinIR \cite{swinir} & 2021 & 27.6 (4.4) & 0.732 (0.116) & 0.253 (0.055) & 0.961 (0.067) & 19.6 (4.0) & 0.667 (0.052) & 0.196 (0.048) & 0.969 (0.012) \\
\multicolumn{1}{c|}{} & U-net \cite{unet} & 2015 & 26.0 (3.3) & 0.674 (0.190) & 0.287 (0.099) & 0.957 (0.083) & 19.5 (2.7) & 0.632 (0.158) & 0.250 (0.084) & 0.939 (0.049) \\
\multicolumn{1}{c|}{} & SwinMR \cite{swinmr} & 2022 & 31.4 (4.3) & 0.855 (0.083) & 0.116 (0.038) & 0.983 (0.045) & 21.5 (4.7) & 0.845 (0.065) & 0.131 (0.041) & 0.982 (0.010) \\
\multicolumn{1}{c|}{5$\%$} & ReconFormer \cite{reconformer} & 2024 & 28.2 (3.2) & 0.673 (0.075) & 0.247 (0.044) & 0.967 (0.057) & 22.1 (3.5) & 0.718 (0.090) & 0.193 (0.034) & 0.973 (0.011) \\
\multicolumn{1}{c|}{} & Video-INR \cite{videoinr} & 2022 & 27.3 (4.2) & 0.800 (0.102) & 0.179 (0.053) & 0.969 (0.069) & 25.1 (3.7) & 0.881 (0.050) & 0.096 (0.026) & 0.980 (0.008) \\
\multicolumn{1}{c|}{} & CycleINR \cite{cycleinr} & 2024 & 24.2 (4.1) & 0.729 (0.136) & 0.245 (0.076) & 0.955 (0.086) & 19.8 (3.1) & 0.800 (0.087) & 0.169 (0.053) & 0.959 (0.040) \\
\multicolumn{1}{c|}{} & SA-INR \cite{sainr} & 2024 & 31.3 (4.3) & 0.826 (0.089) & \textbf{0.131 (0.045)} & 0.979 (0.047) & 25.7 (3.3) & 0.891 (0.056) & \textbf{0.087 (0.029)} & 0.984 (0.009)  \\
\multicolumn{1}{c|}{} & FCB \cite{FCB} & 2025 & 26.4 (4.1) & 0.808 (0.081) & 0.167 (0.063) & 0.961 (0.050) & 22.5 (4.4) & 0.852 (0.062) & 0.134 (0.041) & 0.979 (0.010) \\
\multicolumn{1}{c|}{} & MMR-Mamba \cite{MMR-Mamba} & 2025 & 26.1 (5.5) & 0.829 (0.114) & 0.149 (0.089) & 0.967 (0.068) & 23.3 (6.7) & 0.816 (0.064) & 0.085 (0.041) & 0.987 (0.011) \\
\multicolumn{1}{c|}{} & Our PR-INR & 2025 & \textbf{31.5 (2.4)} & \textbf{0.867 (0.034)} & 0.137 (0.057) & \textbf{0.984 (0.013)} & \textbf{26.4 (4.6)} & \textbf{0.895 (0.064)} & 0.090 (0.048) & \textbf{0.987 (0.011)} \\
\hline
\hline
\multicolumn{1}{c|}{AF} & \multicolumn{1}{c|}{Methods} & \multicolumn{1}{c|}{Years} 
& PSNR$\uparrow$ & SSIM$\uparrow$ & LPIPS$\downarrow$ & NCC$\uparrow$ 
& PSNR$\uparrow$ & SSIM$\uparrow$ & LPIPS$\downarrow$ & NCC$\uparrow$ \\
\hline
\multicolumn{1}{c|}{} & Zero-filled & - & 23.2 (2.9) & 0.535 (0.107) & 0.350 (0.053) & 0.940 (0.033) & 19.7 (2.6) & 0.599 (0.064) & 0.299 (0.025) & 0.945 (0.017) \\
\multicolumn{1}{c|}{} & D5C5 \cite{D5C5} & 2017 & 26.6 (3.1) & 0.635 (0.092) & 0.303 (0.066) & 0.955 (0.067)  & 21.9 (4.3) & 0.794 (0.055) & 0.163 (0.047) & 0.968 (0.016) \\
\multicolumn{1}{c|}{} & MoDL \cite{MoDL} & 2018 & 27.1 (2.6) & 0.706 (0.105) & 0.243 (0.086) & 0.956 (0.088) & 22.1 (4.9) & 0.697 (0.323) & 0.108 (0.050) & 0.971 (0.067) \\
\multicolumn{1}{c|}{} & CycleGAN \cite{Cyclegan} & 2017 & 26.2 (3.2) & 0.716 (0.117) & 0.275 (0.076) & 0.959 (0.065) & 20.2 (3.8) & 0.740 (0.118) & 0.133 (0.035) & 0.967 (0.060) \\
\multicolumn{1}{c|}{} & SwinIR \cite{swinir} & 2021 & 25.7 (2.7) & 0.729 (0.098) & 0.217 (0.040) & 0.965 (0.031) & 23.8 (5.3) & 0.880 (0.057) & 0.092 (0.042) & 0.987 (0.090) \\
\multicolumn{1}{c|}{} & U-net \cite{unet} & 2015 & 25.1 (2.6) & 0.626 (0.180) & 0.285 (0.087) & 0.955 (0.053) & 20.9 (3.5) & 0.654 (0.186) & 0.230 (0.093) & 0.945 (0.034) \\
\multicolumn{1}{c|}{} & SwinMR \cite{swinmr} & 2022 & 27.2 (2.8) & 0.723 (0.084) & 0.211 (0.044) & 0.969 (0.026) & 23.1 (5.6) & 0.874 (0.063) & 0.097 (0.043) & 0.987 (0.009) \\
\multicolumn{1}{c|}{4x} & ReconFormer \cite{reconformer} & 2024 & \textbf{27.6 (2.4)} & 0.737 (0.066) & 0.277 (0.054) & 0.973 (0.037) & 22.6 (2.9) & 0.684 (0.063) & 0.190 (0.026) & 0.980 (0.010) \\
\multicolumn{1}{c|}{} & Video-INR \cite{videoinr} & 2022 & 25.7 (2.8) & 0.731 (0.084) & 0.220 (0.024) & 0.960 (0.056) & 26.5 (3.8) & 0.911 (0.059) & 0.075 (0.037) & 0.987 (0.008) \\
\multicolumn{1}{c|}{} & CycleINR \cite{cycleinr} & 2024 & 22.5 (3.4) & 0.723 (0.096) & 0.241 (0.072) & 0.958 (0.056) & 20.5 (3.3) & 0.821 (0.076) & 0.131 (0.041) & 0.966 (0.050) \\
\multicolumn{1}{c|}{} & SA-INR \cite{sainr} & 2024 & 25.1 (2.8) & 0.729 (0.075) & \textbf{0.185 (0.045)} & \textbf{0.973 (0.026)} & 26.4 (3.3) & 0.884 (0.069) & 0.087 (0.041) & 0.985 (0.009) \\
\multicolumn{1}{c|}{} & FCB \cite{FCB} & 2025 & 25.6 (3.1) & 0.728 (0.071) & 0.225 (0.059) & 0.950 (0.037) & 23.4 (5.6) & 0.874 (0.060) & 0.115 (0.035) & 0.986 (0.007) \\
\multicolumn{1}{c|}{} & MMR-Mamba \cite{MMR-Mamba} & 2025 & 24.8 (6.2) & 0.728 (0.096)  & 0.174 (0.078) & 0.970 (0.036) & 26.7 (5.7) & 0.875 (0.068) & 0.087 (0.045) & \textbf{0.988 (0.011)} \\
\multicolumn{1}{c|}{} & Our PR-INR & 2025 & 27.5 (1.7) & \textbf{0.755 (0.049)} & 0.187 (0.045) & 0.970 (0.009) & \textbf{27.0 (3.4)} & \textbf{0.907 (0.061)} & \textbf{0.087 (0.044)} &  0.987 (0.009) \\

\hline
\multicolumn{1}{c|}{} & Zero-filled & - & 20.2 (2.9) & 0.427 (0.114) & 0.456 (0.059) & 0.886 (0.059) & 17.1 (2.4) & 0.449 (0.101) & 0.442 (0.053) & 0.881 (0.031) \\
\multicolumn{1}{c|}{} & D5C5 \cite{D5C5} & 2017 & 23.2 (3.6) & 0.465 (0.139) & 0.347 (0.043) & 0.822 (0.137) & 17.0 (3.2) & 0.633 (0.090) & 0.296 (0.060) & 0.903 (0.031) \\
\multicolumn{1}{c|}{} & MoDL \cite{MoDL} & 2018 & 24.9 (2.7) & 0.655 (0.124) & 0.284 (0.085) & 0.937 (0.117) & 18.7 (3.8) & 0.607 (0.156) & 0.216 (0.057) & 0.934 (0.068) \\
\multicolumn{1}{c|}{} & CycleGAN \cite{Cyclegan} & 2017 & 24.5 (2.8) & 0.676 (0.116) & 0.297 (0.068) & 0.942 (0.074) & 16.8 (3.4) & 0.604 (0.118) & 0.228 (0.046) & 0.936 (0.060)  \\
\multicolumn{1}{c|}{} & SwinIR \cite{swinir} & 2021 & 24.6 (3.3) & 0.646 (0.102) & 0.280 (0.040) & 0.940 (0.051) & 19.1 (3.8) & 0.743 (0.095) & 0.241 (0.059) & 0.956 (0.023) \\
\multicolumn{1}{c|}{} & U-net \cite{unet} & 2015 & 22.9 (2.8) & 0.549 (0.166) & 0.333 (0.082) & 0.924 (0.073) & 18.3 (3.0) & 0.592 (0.159) & 0.283 (0.078) & 0.922 (0.049) \\
\multicolumn{1}{c|}{} & SwinMR \cite{swinmr} & 2022 & 25.8 (3.5) & 0.682 (0.096) & 0.238 (0.048) & 0.949 (0.044) & 18.5 (4.5) & 0.711 (0.086) & 0.191 (0.055) & 0.964 (0.019) \\
\multicolumn{1}{c|}{8x} & ReconFormer \cite{reconformer} & 2024 & 25.9 (2.5) & 0.578 (0.073) & 0.320 (0.053) & 0.959 (0.056) & 20.1 (2.8) & 0.604 (0.084) & 0.272 (0.036) & 0.956 (0.019) \\
\multicolumn{1}{c|}{} & Video-INR \cite{videoinr} & 2022 & 24.9 (3.6) & 0.663 (0.098) & 0.264 (0.047) & 0.924 (0.064) & 24.3 (3.0) & 0.875 (0.068) & \textbf{0.099 (0.047)} & 0.978 (0.013) \\
\multicolumn{1}{c|}{} & CycleINR \cite{cycleinr} & 2024 & 22.9 (2.8) & 0.707 (0.094) & 0.240 (0.070) & 0.953 (0.056) & 18.3 (2.9) & 0.755 (0.093) & 0.200 (0.061) & 0.943 (0.046) \\
\multicolumn{1}{c|}{} & SA-INR \cite{sainr} & 2024 & 26.1 (3.0) & 0.667 (0.089) & 0.231 (0.051) & 0.947 (0.050) & 22.7 (2.4) & 0.824 (0.078) & 0.148 (0.047) & 0.966 (0.018) \\
\multicolumn{1}{c|}{} & FCB \cite{FCB} & 2025 & 25.1 (3.5) & 0.705 (0.073) & 0.196 (0.066) & 0.953 (0.045) & 21.6 (3.0) & 0.778 (0.086) & 0.208 (0.059) & 0.955 (0.021) \\
\multicolumn{1}{c|}{} & MMR-Mamba \cite{MMR-Mamba} & 2025 & 23.1 (5.5) & 0.666 (0.099) & 0.225 (0.085) & 0.955 (0.064) & 20.6 (6.1) & 0.789 (0.086) & 0.122 (0.059) & 0.977 (0.019) \\
\multicolumn{1}{c|}{} & Our PR-INR & 2025 & \textbf{26.4 (2.0)} & \textbf{0.721 (0.037)} & \textbf{0.201 (0.040)} & \textbf{0.957 (0.011)} & \textbf{24.5 (4.1)} & \textbf{0.879 (0.066)} & 0.105 (0.060) & \textbf{0.980 (0.017)} \\
\hline
\end{tabular}}
}
\end{table*}

\begin{table}[t]
\centering
{
\caption{Quantitative comparison with state-of-the-art MRI reconstruction methods on the BraTS dataset under motion corruption and acceleration.}
\label{tab:brats}
\resizebox{\columnwidth}{!}{
\begin{tabular}{ll|l|rrrr}
\hline
\multicolumn{3}{c|}{} & \multicolumn{4}{c}{BraTS} \\ 
\hline
\multicolumn{1}{c|}{DM} & \multicolumn{1}{c|}{Methods} & \multicolumn{1}{c|}{Years} & PSNR$\uparrow$ & SSIM$\uparrow$ & LPIPS$\downarrow$ & NCC$\uparrow$ \\
\hline\hline

\multicolumn{1}{c|}{} & Zero-filled & - & 24.3 (1.3) & 0.754 (0.025) & 0.220 (0.016) & 0.935 (0.018) \\
\multicolumn{1}{c|}{} & D5C5 \cite{D5C5} & 2017 & 26.0 (1.3) & 0.853 (0.016) & 0.179 (0.014) & 0.957 (0.011) \\
\multicolumn{1}{c|}{} & MoDL \cite{MoDL} & 2018 & 28.2 (1.6) & 0.851 (0.031) & 0.200 (0.029) & 0.979 (0.006) \\
\multicolumn{1}{c|}{} & CycleGAN \cite{Cyclegan} & 2017 & 24.8 (2.4) & 0.766 (0.054) & 0.125 (0.019) & 0.972 (0.008) \\
\multicolumn{1}{c|}{} & SwinIR \cite{swinir} & 2021 & 30.3 (1.8) & 0.940 (0.009) & 0.063 (0.008) & 0.987 (0.003) \\
\multicolumn{1}{c|}{} & U-net \cite{unet} & 2015 & 26.0 (1.6) & 0.835 (0.015) & 0.177 (0.010) & 0.957 (0.010) \\
\multicolumn{1}{c|}{} & SwinMR \cite{swinmr} & 2022 & 30.6 (2.2) & 0.950 (0.005) & 0.053 (0.008) & 0.990 (0.003) \\
\multicolumn{1}{c|}{3$\%$} & ReconFormer \cite{reconformer} & 2024 & 28.7 (1.3) & 0.806 (0.020) & 0.135 (0.015) & 0.980 (0.005) \\
\multicolumn{1}{c|}{} & Video-INR \cite{videoinr} & 2022 & 31.5 (1.5) & 0.961 (0.005) & 0.044 (0.007) & 0.989 (0.002) \\
\multicolumn{1}{c|}{} & CycleINR \cite{cycleinr} & 2024 & 27.2 (1.8) & 0.925 (0.010) & 0.082 (0.009) & 0.972 (0.007) \\
\multicolumn{1}{c|}{} & SA-INR \cite{sainr} & 2024 & 32.1 (1.9) & 0.958 (0.004) & \textbf{0.031} (0.007) & 0.983 (0.002) \\
\multicolumn{1}{c|}{} & FCB \cite{FCB} & 2025 & 28.0 (2.2) & 0.925 (0.009) & 0.089 (0.010) & 0.982 (0.004) \\
\multicolumn{1}{c|}{} & MMR-Mamba \cite{MMR-Mamba} & 2025 & 30.5 (2.6) & 0.937 (0.013) & 0.054 (0.008) & 0.988 (0.003) \\
\multicolumn{1}{c|}{} & Our PR-INR & 2025 & \textbf{32.1 (1.2)} & \textbf{0.962 (0.004)} & 0.046 (0.007) & \textbf{0.991 (0.003)} \\

\hline
\multicolumn{1}{c|}{} & Zero-filled & - & 21.7 (1.2) & 0.544 (0.042) & 0.332 (0.025) & 0.897 (0.027) \\
\multicolumn{1}{c|}{} & D5C5 \cite{D5C5} & 2017 & 22.1 (1.5) & 0.777 (0.023) & 0.238 (0.019) & 0.927 (0.018) \\
\multicolumn{1}{c|}{} & MoDL \cite{MoDL} & 2018 & 26.0 (1.6) & 0.789 (0.035) & 0.239 (0.032) & 0.964 (0.010) \\
\multicolumn{1}{c|}{} & CycleGAN \cite{Cyclegan} & 2017 & 24.8 (2.4) & 0.766 (0.054) & 0.125 (0.019) & 0.972 (0.008) \\
\multicolumn{1}{c|}{} & SwinIR \cite{swinir} & 2021 & 27.6 (2.0) & 0.742 (0.018) & 0.115 (0.012) & 0.974 (0.006) \\
\multicolumn{1}{c|}{} & U-net \cite{unet} & 2015 & 25.0 (1.4) & 0.815 (0.020) & 0.190 (0.012) & 0.947 (0.012) \\
\multicolumn{1}{c|}{} & SwinMR \cite{swinmr} & 2022 & 27.2 (2.3) & 0.900 (0.011) & 0.101 (0.011) & 0.978 (0.006) \\
\multicolumn{1}{c|}{5$\%$} & ReconFormer \cite{reconformer} & 2024 & 26.5 (1.6) & 0.678 (0.033) & 0.189 (0.019) & 0.968 (0.008) \\
\multicolumn{1}{c|}{} & Video-INR \cite{videoinr} & 2022 & 28.7 (1.7) & 0.933 (0.008) & 0.072 (0.009) & 0.981 (0.004) \\
\multicolumn{1}{c|}{} & CycleINR \cite{cycleinr} & 2024 & 26.3 (1.5) & 0.899 (0.010) & 0.103 (0.011) & 0.966 (0.006) \\
\multicolumn{1}{c|}{} & SA-INR \cite{sainr} & 2024 & 29.5 (1.7) & 0.934 (0.007) & 0.069 (0.010) & 0.985 (0.004) \\
\multicolumn{1}{c|}{} & FCB \cite{FCB} & 2025 & 27.5 (2.6) & 0.882 (0.014) & 0.115 (0.012) & 0.970 (0.008) \\
\multicolumn{1}{c|}{} & MMR-Mamba \cite{MMR-Mamba} & 2025 & 27.6 (2.6) & 0.911 (0.016) & 0.075 (0.010) & 0.982 (0.004) \\
\multicolumn{1}{c|}{} & Our PR-INR & 2025 & \textbf{30.0 (1.1)} & \textbf{0.942 (0.004)} & \textbf{0.067 (0.009)} & \textbf{0.985 (0.004)} \\

\hline
\multicolumn{1}{c|}{AF} & \multicolumn{1}{c|}{Methods} & \multicolumn{1}{c|}{Years} & PSNR$\uparrow$ & SSIM$\uparrow$ & LPIPS$\downarrow$ & NCC$\uparrow$ \\
\hline
\multicolumn{1}{c|}{} & Zero-filled & - & 27.3 (1.2) & 0.686 (0.023) & 0.198 (0.021) & 0.978 (0.005) \\
\multicolumn{1}{c|}{} & D5C5 \cite{D5C5} & 2017 & 30.4 (1.3) & 0.937 (0.005) & 0.080 (0.009) & 0.984 (0.004) \\
\multicolumn{1}{c|}{} & MoDL \cite{MoDL} & 2018 & 30.9 (1.6) & 0.930 (0.054) & 0.030 (0.005) & 0.993 (0.001) \\
\multicolumn{1}{c|}{} & CycleGAN \cite{Cyclegan} & 2017 & 26.6 (3.0) & 0.909 (0.027) & 0.080 (0.014) & 0.980 (0.004) \\
\multicolumn{1}{c|}{} & SwinIR \cite{swinir} & 2021 & 31.1 (3.0) & 0.895 (0.013) & 0.035 (0.007) & 0.994 (0.001) \\
\multicolumn{1}{c|}{} & U-net \cite{unet} & 2015 & 26.8 (3.8) & 0.897 (0.023) & 0.104 (0.011) & 0.982 (0.003) \\
\multicolumn{1}{c|}{} & SwinMR \cite{swinmr} & 2022 & 33.9 (2.6) & 0.974 (0.005) & 0.022 (0.005) & 0.996 (0.001) \\
\multicolumn{1}{c|}{4x} & ReconFormer \cite{reconformer} & 2024 & 30.1 (1.6) & 0.720 (0.022) & 0.118 (0.014) & 0.991 (0.002) \\
\multicolumn{1}{c|}{} & Video-INR \cite{videoinr} & 2022 & 32.9 (2.7) & 0.980 (0.003) & 0.021 (0.004) & 0.996 (0.001) \\
\multicolumn{1}{c|}{} & CycleINR \cite{cycleinr} & 2024 & 28.7 (1.4) & 0.941 (0.007) & 0.062 (0.009) & 0.967 (0.008) \\
\multicolumn{1}{c|}{} & SA-INR \cite{sainr} & 2024 & 32.5 (2.3) & 0.980 (0.002) & \textbf{0.016 (0.004)} & 0.996 (0.001) \\
\multicolumn{1}{c|}{} & FCB \cite{FCB} & 2025 & 30.0 (2.2) & 0.952 (0.005) & 0.053 (0.007) & 0.989 (0.003) \\
\multicolumn{1}{c|}{} & MMR-Mamba \cite{MMR-Mamba} & 2025 & 28.7 (2.6) & 0.949 (0.012) & 0.036 (0.006) & 0.992 (0.002) \\
\multicolumn{1}{c|}{} & Our PR-INR & 2025 & \textbf{34.1 (1.1)} & \textbf{0.984 (0.003)} & 0.020 (0.006) & \textbf{0.996 (0.002)} \\

\hline
\multicolumn{1}{c|}{} & Zero-filled & - & 21.6 (1.6) & 0.599 (0.031) & 0.295 (0.023) & 0.929 (0.015) \\
\multicolumn{1}{c|}{} & D5C5 \cite{D5C5} & 2017 & 24.3 (2.3) & 0.863 (0.014) & 0.131 (0.014) & 0.953 (0.011) \\
\multicolumn{1}{c|}{} & MoDL \cite{MoDL} & 2018 & 27.5 (2.2) & 0.907 (0.034) & 0.056 (0.009) & 0.983 (0.004) \\
\multicolumn{1}{c|}{} & CycleGAN \cite{Cyclegan} & 2017 & 23.5 (2.4) & 0.790 (0.044) & 0.111 (0.016) & 0.968 (0.006) \\
\multicolumn{1}{c|}{} & SwinIR \cite{swinir} & 2021 & 24.1 (2.1) & 0.756 (0.020) & 0.162 (0.015) & 0.954 (0.010) \\
\multicolumn{1}{c|}{} & U-net \cite{unet} & 2015 & 23.9 (2.5) & 0.813 (0.024) & 0.180 (0.012) & 0.951 (0.010) \\
\multicolumn{1}{c|}{} & SwinMR \cite{swinmr} & 2022 & 26.8 (2.8) & 0.898 (0.015) & 0.079 (0.010) & 0.976 (0.005) \\
\multicolumn{1}{c|}{8x} & ReconFormer \cite{reconformer} & 2024 & 25.8 (2.4) & 0.683 (0.036) & 0.138 (0.014) & 0.980 (0.004) \\
\multicolumn{1}{c|}{} & Video-INR \cite{videoinr} & 2022 & 27.8 (1.8) & 0.931 (0.009) & 0.047 (0.009) & 0.983 (0.004) \\
\multicolumn{1}{c|}{} & CycleINR \cite{cycleinr} & 2024 & 27.1 (1.7) & 0.915 (0.011) & 0.084 (0.011) & 0.971 (0.006) \\
\multicolumn{1}{c|}{} & SA-INR \cite{sainr} & 2024 & 27.3 (1.6) & 0.931 (0.008) & \textbf{0.043 (0.009)} & 0.984 (0.003) \\
\multicolumn{1}{c|}{} & FCB \cite{FCB} & 2025 & 26.7 (2.3) & 0.859 (0.015) & 0.140 (0.013) & 0.953 (0.011) \\
\multicolumn{1}{c|}{} & MMR-Mamba \cite{MMR-Mamba} & 2025 & 26.3 (3.4) & 0.902 (0.021) & 0.057 (0.011) & 0.982 (0.004) \\
\multicolumn{1}{c|}{} & Our PR-INR & 2025 & \textbf{29.5 (1.2)} & \textbf{0.935 (0.008)} & 0.045 (0.010) & \textbf{0.984 (0.005)} \\
\hline
\end{tabular}}
}
\end{table}

To train the conditioned INR with robust supervision, we adopt a dual-domain residual loss that constrains the predicted refinement $r(x,y)$ in both image and frequency domains. The loss is defined as:
\begin{equation}
\mathcal{L}_{\text{IDR}} = \left\| {x}_{\text{IDR}} - \tilde{x}_0 \right\|_2 \;+\; \left\| w \odot \left( \mathcal{F}({x}_{\text{IDR}}) - \mathcal{F}(\tilde{x}_0) \right) \right\|_2,
\end{equation}
where $\tilde{x}_0$ is the data-consistent image reconstructed from diffusion outputs, $\mathcal{F}(\cdot)$ is the Fourier transform, and $w$ is a frequency weighting mask that emphasizes high-frequency components. This residual supervision encourages the INR to selectively refine under-constrained spectral regions while preserving the reliable low-frequency components already restored by the diffusion and data consistency modules. 
{Specifically, the weighting mask $w$ is implemented as a fixed two-dimensional radial map in k-space, where the central low-frequency region is softly down-weighted and the outer high-frequency region is progressively emphasized. The weight at each frequency location is determined by its normalized distance from the k-space center. This mask provides stronger supervision for high-frequency residuals to guide the INR refinement in under-constrained regions. All image-domain signals are normalized to the range [0,1] before loss computation and the residual energy is preserved across domains, ensuring that both loss terms contribute in a balanced manner.} This residual supervision encourages the INR to selectively refine under-constrained spectral regions while preserving the reliable low-frequency components already restored by the diffusion and data consistency modules.

To further enhance generalization under diverse undersampling patterns, we employ a mask-randomized pretraining strategy, followed by fine-tuning under the target sampling pattern. 
{Specifically, during pretraining, four types of undersampling masks are randomly applied, including 1D Gaussian, 2D Gaussian, radial, and spiral patterns, with acceleration factors of 4× and 8×. This setting exposes the network to various aliasing artifacts and sampling geometries, encouraging the diffusion and INR modules to learn structure-aware priors rather than pattern-specific mappings.}

The combination of data-consistency correction for low-frequency spectrum, residual INR for selective high-frequency refinement, and the Fourier-inspired representation capacity of SIREN allows us to significantly reduce diffusion's hallucination space:
\begin{equation}
\operatorname{Var}({x}_{\text{IDR}} \mid y) \ll \operatorname{Var}({x}_{\text{MAD}} \mid y),
\end{equation}
leading to reconstructions that are both faithful and rich in details. The proof details are in the Section \ref{appendix} Appendix.

\subsubsection{Voxel Continuous-aware Representation Module for Volumetric Anisotropy}
While the conditioned INR refines slice-wise details, achieving true slice-to-volume consistency requires a volumetric neural representation. To this end, we introduce a Voxel Continuous-aware Representation (VCR) Module $f_{\theta_3}(x,y,z)$ that directly maps spatial coordinates to intensity values (Fig. \ref{fig3}(c)). {The VCR module is constructed as a 3D conditional SIREN-based implicit neural representation with five hidden layers and 256 hidden units per layer, using sine activations and frequency scaling parameters $\omega_0^{(1)} = 30$ and $\omega_0^{(h)} = 30$. To explicitly capture inter-slice dependencies, a ConvLSTM block with depth 4, hidden size 128, and kernel size 3×3 is integrated into the network, propagating spatial-temporal context across adjacent slices.}

Given the refined slices $\{f_i(x,y)\}_{i=1}^M$ from previous stages located at $z=z_i$, we aim to fit $f_{\theta_3}$ such that it reproduces these slices when queried at the corresponding planes, while maintaining spatial smoothness across $z$.

To facilitate this, multi-scale features are first extracted from each slice using a 2D encoder. These features are then fed into a bidirectional ConvLSTM to capture inter-slice correlations in both forward and backward directions. The resulting hidden states are aggregated into a volumetric latent representation $\mathbf{c}_2(x,y,z)$, which encodes both local texture and contextual information across adjacent slices. The volumetric INR $f_{\theta_3}$ then takes $(x,y,z)$ and the aggregated feature $\mathbf{c}_2(x,y,z)$ as input.

The learning objective is defined as:
\begin{equation}
\label{eq:vol_loss}
\mathcal{L}_{\text{VCR}}=\;\frac{1}{M} \sum_{i=1}^{M} \int \Bigl|f_{\theta_3}\bigl(x,y,z_i;\mathbf{c}_2(x,y,z)\bigr)\;-\;f_i(x,y)\Bigr|^2 \,dx\,dy,
\end{equation}
where $f_i(x,y)$ denotes the observed slice at location $z=z_i$.
\textcolor[RGB]{0,0,0}{Under thick-slice acquisition, $f_i(x,y)$ represents an anisotropically blurred observation over a finite through-plane extent, rather than a voxel-resolved ground-truth slice at an exact $z$ location. Accordingly, Eq.~(\ref{eq:vol_loss}) is formulated as a soft consistency constraint that regularizes the shared continuous volumetric representation using the available slice observations, rather than enforcing exact voxel-wise equality.}

By fitting $f_{\theta_3}$ to each available slice in this weakly supervised manner, we encourage the network to build a spatially smooth and coherent representation along the slice direction. The continuity of $f_{\theta_3}(x,y,z)$ naturally enforces smooth variation along the $z$-axis, implying:
\begin{equation}
\bigl|\,f_{\theta_3}(x,y,z+\Delta z) - f_{\theta_3}(x,y,z)\bigr|\;\le\;C\,|\Delta z|,
\end{equation}
where $C$ denotes a Lipschitz constant that characterizes the spatial smoothness of the volumetric representation along the slice axis. \textcolor[RGB]{0,0,0}{Importantly, this smoothness constraint complements the VCR objective without assuming exact slice-to-volume correspondence under thick-slice sampling.} In practice, sinusoidal activations and Fourier feature encodings can further enhance the capacity to capture high-frequency anatomical structures.

By merging slice-wise features into a continuous volumetric function, we not only retain the high-frequency details recovered in earlier stages but also fill inter-slice gaps for volumetric reconstruction. This design accommodates uneven slice spacing or anisotropic scans, as $f_{\theta_3}$ can be queried at arbitrary $z$ coordinates to generate intermediate slices or re-slice the volume in different orientations.

\subsection{Overall Loss Function}

The overall loss of the proposed PR-INR framework integrates the objectives of diffusion denoising, dual-domain residual supervision, and volumetric smoothing into a unified formulation:

\begin{equation}
\mathcal{L}_{\text{total}} =  \lambda_1 \mathcal{L}_{\text{MAD}} + \lambda_2 \mathcal{L}_{\text{IDR}} + \lambda_3 \mathcal{L}_{\text{VCR}}
\end{equation}

where $\mathcal{L}_{\text{MAD}}$ is the diffusion-based denoising loss, $\mathcal{L}_{\text{IDR}}$ represents the image-domain and k-space-domain residual losses computed slice-wise for the IDR module, and $\mathcal{L}_{\text{VCR}}$ enforces inter-slice continuity and volumetric smoothness. 
{The weights are set to $\lambda_1 = 0.5$, $\lambda_2 = 1.0$, and $\lambda_3 = 0.3$, which balance the contributions of global structural regularization, slice-level detail refinement, and volumetric consistency. These coefficients were determined through a grid search on the validation set, where different combinations were evaluated to identify the configuration that achieved the best trade-off between edge sharpness and volumetric smoothness. We used the same weighting configuration for all datasets, which consistently yielded stable performance and demonstrated strong cross-dataset generalizability.} This balanced weighting enables the PR-INR framework to achieve stable and high-fidelity slice-to-volume reconstruction across different anatomical regions.

\begin{table}[t]
\centering
\caption{Performance comparison of different methods under scales.}
\resizebox{\columnwidth}{!}{
\begin{tabular}{c|cc|cccc}
\hline
\multicolumn{1}{c|}{\textbf{Method}} & \multicolumn{1}{c}{\textbf{PSNR}$\uparrow$} & \multicolumn{1}{c|}{\textbf{SSIM}$\uparrow$} & \multicolumn{1}{c}{\textbf{DSC (ET)}$\uparrow$} & \multicolumn{1}{c}{\textbf{DSC (TC)$\uparrow$}} & \multicolumn{1}{c}{\textbf{DSC (WT)$\uparrow$}} & \multicolumn{1}{c}{\textbf{DSC (AVG)$\uparrow$}} \\ \hline
\multicolumn{1}{c|}{Full-sample} & - & - & 71.7 (12.5) & 87.0 (5.6) & 83.1 (9.2) & 80.6 (6.5) \\ \hline
Repeat & 30.7 (4.2) & 0.902 (0.077) & 64.8 (12.6) & 85.6 (5.2) & 80.3 (8.7) & 76.9 (8.8) \\
Trilinear & 31.2 (4.1) & 0.910 (0.072) & 66.6 (12.1) & 85.4 (5.2) & 80.6 (8.3) & 77.6 (8.0) \\
Video-INR \cite{videoinr} & 32.8 (3.9) & 0.930 (0.060) & 67.4 (11.9) & 86.2 (5.0) & 81.2 (8.0) & 78.3 (7.6) \\
% CycleINR \cite{cycleinr} & 30.2 (3.7) & 0.915 (0.057) & 65.0 (11.5) & 84.4 (4.8) & 76.7 (7.7) & 78.7 (7.4) \\
SA-INR \cite{sainr} & 33.7 (3.6) & 0.938 (0.054) & 68.7 (11.2) & 86.6 (4.7) & 82.1 (7.5) & 79.1 (7.2) \\
\textbf{Our PR-INR} & \textbf{34.4 (3.3)} & \textbf{0.941 (0.051)} & \textbf{69.4 (11.0)} & \textbf{87.0 (4.6)} & \textbf{82.6 (7.2)} & \textbf{79.7 (7.0)} \\ \hline
\end{tabular}
}
\label{compar_3D}
\end{table}

\begin{table}[t]
\centering
\caption{Joint ablation study on FastMRI dataset and HCP 1200 dataset. $\checkmark$ indicates the module is enabled. Best results are in bold.}
\label{tab:ablation_final}
\resizebox{\columnwidth}{!}{
\begin{tabular}{c|p{4mm}p{4mm}p{4mm}p{4mm}|cccc}
\hline
\multicolumn{9}{c}{FastMRI Dataset}\\
\hline
\textbf{Type} & \textbf{Base.} & \textbf{MAD} & \textbf{IDR} & \textbf{VCR} 
& \textbf{PSNR$\uparrow$} & \textbf{SSIM$\uparrow$} & \textbf{LPIPS$\downarrow$} & \textbf{NCC$\uparrow$} \\
\hline
\multirow{4}{*}{$3\%$} 
& \checkmark & x & x & x & 30.0 (3.6) & 0.853 (0.082) & 0.149 (0.075) & 0.965 (0.053) \\
& \checkmark & \checkmark & x & x & 31.7 (4.2) & 0.863 (0.088) & 0.120 (0.057) & 0.974 (0.054) \\
& \checkmark & \checkmark & \checkmark & x & 32.6 (3.9) & 0.874 (0.065) & 0.092 (0.038) & 0.988 (0.018) \\
& \checkmark & \checkmark & \checkmark & \checkmark & \textbf{33.3 (2.3)} & \textbf{0.919 (0.035)} & \textbf{0.063 (0.029)} & \textbf{0.990 (0.007)} \\

\hline
\multirow{4}{*}{$5\%$} 
& \checkmark & x & x & x & 28.3 (3.4) & 0.775 (0.129) & 0.184 (0.093) & 0.940 (0.110) \\
& \checkmark & \checkmark & x & x & 29.8 (3.7) & 0.805 (0.131) & 0.158 (0.071) & 0.968 (0.050) \\
& \checkmark & \checkmark & \checkmark & x & 30.9 (3.9) & 0.846 (0.092) & 0.151 (0.072) & 0.971 (0.050) \\
& \checkmark & \checkmark & \checkmark & \checkmark & \textbf{31.5 (2.4)} & \textbf{0.867 (0.034)} & \textbf{0.137 (0.057)} & \textbf{0.984 (0.013)} \\
\hline

\multirow{4}{*}{4x} 
& \checkmark & x & x & x & 25.2 (3.2) & 0.716 (0.082) & 0.192 (0.083) & 0.954 (0.066) \\
& \checkmark & \checkmark & x & x & 26.1 (3.1) & 0.749 (0.083) & 0.176 (0.083) & 0.958 (0.048) \\
& \checkmark & \checkmark & \checkmark & x & 26.7 (2.9) & 0.764 (0.080) & 0.185 (0.085) & 0.965 (0.051) \\
& \checkmark & \checkmark & \checkmark & \checkmark & \textbf{27.5 (1.7)} & \textbf{0.755 (0.049)} & \textbf{0.187 (0.045)} & \textbf{0.970 (0.009)} \\
\hline

\multirow{4}{*}{8x} 
& \checkmark & x & x & x & 23.9 (3.1) & 0.645 (0.096) & 0.228 (0.087) & 0.928 (0.100) \\
& \checkmark & \checkmark & x & x & 25.0 (3.1) & 0.681 (0.088) & 0.203 (0.085) & 0.940 (0.083) \\
& \checkmark & \checkmark & \checkmark & x & 25.5 (3.2) & 0.714 (0.087) & 0.198 (0.085) & 0.939 (0.078) \\
& \checkmark & \checkmark & \checkmark & \checkmark & \textbf{26.4 (2.0)} & \textbf{0.721 (0.037)} & \textbf{0.201 (0.040)} & \textbf{0.957 (0.011)} \\

\hline
\hline
\multicolumn{9}{c}{HCP 1200 Dataset}\\
\hline
\textbf{Type} & \textbf{Base.} & \textbf{MAD} & \textbf{IDR} & \textbf{VCR} 
& \textbf{PSNR$\uparrow$} & \textbf{SSIM$\uparrow$} & \textbf{LPIPS$\downarrow$} & \textbf{NCC$\uparrow$} \\
\hline
\multirow{4}{*}{$3\%$} 
& \checkmark & x & x & x & 24.3 (3.7) & 0.820 (0.071) & 0.159 (0.045) & 0.973 (0.010) \\
& \checkmark & \checkmark & x & x & 25.2 (4.1) & 0.865 (0.076) & 0.103 (0.041) & 0.986 (0.009) \\
& \checkmark & \checkmark & \checkmark & x & 26.5 (3.9) & 0.874 (0.065) & \textbf{0.092 (0.038)} & \textbf{0.988 (0.008)} \\
& \checkmark & \checkmark & \checkmark & \checkmark & \textbf{27.0 (3.3)} & \textbf{0.904 (0.064)} & 0.093 (0.009) & 0.985 (0.009) \\

\hline
\multirow{4}{*}{$5\%$} 
& \checkmark & x & x & x & 21.8 (3.6) & 0.720 (0.078) & 0.227 (0.065) & 0.954 (0.020) \\
& \checkmark & \checkmark & x & x & 24.2 (4.3) & 0.805 (0.062) & 0.125 (0.052) & 0.983 (0.011) \\
& \checkmark & \checkmark & \checkmark & x & 25.5 (3.9) & 0.782 (0.091) & 0.129 (0.053) & 0.982 (0.011) \\
& \checkmark & \checkmark & \checkmark & \checkmark & \textbf{26.4 (4.6)} & \textbf{0.895 (0.064)} & \textbf{0.090 (0.048)} & \textbf{0.987 (0.011)} \\
\hline

\multirow{4}{*}{4x} 
& \checkmark & x & x & x & 24.6 (4.1) & 0.792 (0.055) & 0.142 (0.043) & 0.978 (0.010) \\
& \checkmark & \checkmark & x & x & 25.6 (4.3) & 0.845 (0.053) & 0.125 (0.045) & 0.983 (0.012) \\
& \checkmark & \checkmark & \checkmark & x & 26.3 (4.6) & 0.857 (0.081) & 0.115 (0.045) & 0.986 (0.010) \\
& \checkmark & \checkmark & \checkmark & \checkmark & \textbf{27.0 (3.4)} & \textbf{0.907 (0.061)} & \textbf{0.087 (0.044)} & \textbf{0.987 (0.009)} \\
\hline

\multirow{4}{*}{8x} 
& \checkmark & x & x & x & 21.5 (3.5) & 0.691 (0.088) & 0.262 (0.062) & 0.949 (0.033) \\
& \checkmark & \checkmark & x & x & 23.4 (5.1) & 0.811 (0.065) & 0.178 (0.050) & 0.970 (0.010) \\
& \checkmark & \checkmark & \checkmark & x & 23.1 (4.1) & 0.819 (0.084) & 0.160 (0.062) & 0.973 (0.020) \\
& \checkmark & \checkmark & \checkmark & \checkmark & \textbf{24.0 (4.6)} & \textbf{0.849 (0.076)} & \textbf{0.115 (0.060)} & \textbf{0.980 (0.017)} \\
\hline
\end{tabular}}
\end{table}

\begin{table}[t]
\centering
\caption{\textcolor[RGB]{0,0,0}{Quantitative comparison under different loss weight configurations for different sampling ratios.}}
\label{tab:loss_metrics}
\resizebox{\columnwidth}{!}{
{\color[RGB]{0,0,0}
\begin{tabular}{c|ccc|cccc}
\hline
Type & $\lambda_1$ & $\lambda_2$ & $\lambda_3$ &
PSNR (dB)$\uparrow$ & SSIM$\uparrow$ & LPIPS$\downarrow$ & NCC$\uparrow$ \\
\hline
\multirow{7}{*}{5\%}
& \textbf{0.5} & \textbf{1.0} & \textbf{0.3} &
\textbf{31.3 $\pm$ 2.4} & \textbf{0.868 $\pm$ 0.034} &
0.136 $\pm$ 0.057 & \textbf{0.984 $\pm$ 0.013} \\  % Default
& 0.3 & 1.0 & 0.3 &
30.7 $\pm$ 2.7 & 0.855 $\pm$ 0.038 &
0.150 $\pm$ 0.061 & 0.979 $\pm$ 0.016 \\  % MAD↓
& 0.5 & 2.0 & 0.3 &
31.0 $\pm$ 2.6 & 0.860 $\pm$ 0.036 &
\textbf{0.133 $\pm$ 0.059} & 0.981 $\pm$ 0.015 \\  % IDR↑
& 0.5 & 1.0 & 0.1 &
30.6 $\pm$ 2.8 & 0.853 $\pm$ 0.040 &
0.151 $\pm$ 0.062 & 0.975 $\pm$ 0018 \\  % VCR↓
& 0.3 & 1.0 & 0.1 &
29.9 $\pm$ 3.0 & 0.842 $\pm$ 0.043 &
0.162 $\pm$ 0.066 & 0.968 $\pm$ 0.021 \\  % MAD↓ + VCR↓
& 1.0 & 0.5 & 0.3 &
30.4 $\pm$ 2.9 & 0.850 $\pm$ 0.041 &
0.156 $\pm$ 0.064 & 0.977 $\pm$ 0.018 \\  % MAD↑ + IDR↓
& 0.5 & 2.0 & 0.5 &
30.5 $\pm$ 2.9 & 0.852 $\pm$ 0.041 &
0.134 $\pm$ 0.065 & 0.976 $\pm$ 0.019 \\  % IDR↑ + VCR↑
\hline

\multirow{7}{*}{8$\times$}
& \textbf{0.5} & \textbf{1.0} & \textbf{0.3} &
\textbf{26.3 $\pm$ 2.0} & \textbf{0.724 $\pm$ 0.037} &
\textbf{0.199 $\pm$ 0.040} & \textbf{0.958 $\pm$ 0.011} \\  % Default
& 0.3 & 1.0 & 0.3 &
25.7 $\pm$ 2.2 & 0.711 $\pm$ 0.040 &
0.213 $\pm$ 0.045 & 0.952 $\pm$ 0.013 \\  % MAD↓
& 0.5 & 2.0 & 0.3 &
26.1 $\pm$ 2.1 & 0.716 $\pm$ 0.039 &
0.206 $\pm$ 0.043 & 0.955 $\pm$ 0.012 \\  % IDR↑
& 0.5 & 1.0 & 0.1 &
25.7 $\pm$ 2.3 & 0.708 $\pm$ 0.041 &
0.214 $\pm$ 0.046 & 0.946 $\pm$ 0.015 \\  % VCR↓
& 0.3 & 1.0 & 0.1 &
25.1 $\pm$ 2.5 & 0.696 $\pm$ 0.045 &
0.228 $\pm$ 0.052 & 0.938 $\pm$ 0.018 \\  % MAD↓ + VCR↓
& 1.0 & 0.5 & 0.3 &
25.5 $\pm$ 2.4 & 0.703 $\pm$ 0.043 &
0.220 $\pm$ 0.050 & 0.949 $\pm$ 0.015 \\  % MAD↑ + IDR↓
& 0.5 & 2.0 & 0.5 &
25.6 $\pm$ 2.4 & 0.704 $\pm$ 0.043 &
0.208 $\pm$ 0.051 & 0.948 $\pm$ 0.016 \\  % IDR↑ + VCR↑
\hline
\end{tabular}
}}
\end{table}

\section{Experiment}
\subsection{Dataset and Preprocessing}
In this study, we use the FastMRI brain dataset \cite{fastmri} and HCP1200 \cite{hcp} to evaluate the reconstruction performance, the BRATS 2021 dataset \cite{brats2021} to assess its capability in 3D continuous reconstruction, and the CMRxRecon 2024 \cite{cmrxrecon} as well as the FastMRI knee dataset \cite{fastmri} to validate its generalization ability across anatomies. The FastMRI brain dataset comprises 7,002 fully sampled brain MR scans acquired using 1.5T and 3T scanners. The HCP1200 dataset provides ultra-high-resolution 7T brain MRIs from 184 healthy participants. The BRATS 2021 dataset includes MRI scans of 2,040 patients with brain tumors. Pixel-wise annotations cover enhancing tumor (ET), peritumoral edema (ED), and necrotic core (NCR). The tumor core (TC) comprises ET and NCR, while the whole tumor (WT) includes TC and ED. The CMRxRecon 2024 dataset contains multi-view, multi-slice, and multi-coil cardiac MRI k-space data from 300 healthy subjects. It includes cine acquisitions in short-axis and long-axis views for dynamic cardiac imaging and reconstruction tasks. The FastMRI knee dataset includes more than 1,500 fully sampled knee MRIs obtained on 3 and 1.5T scanners.

Preprocessing of multi-coil and complex data: For multi-coil $k$-space acquisitions, we first apply an inverse Fourier transform to each coil channel and then perform root-sum-of-squares (RSS) reconstruction \cite{rss} in the image domain to obtain single-coil magnitude images. For complex-valued $k$-space inputs, we separate the real and imaginary components into two channels. This allows the network to operate in the real domain without losing phase information.

Motion simulation and undersampling strategy: To evaluate the robustness of our framework against motion artifacts and undersampled acquisition, we simulate k-space acquisition line by line. For motion corruption, rigid motion is applied independently at each phase-encoding line, including random in-plane displacement (3–5\% ) of the field of view (FOV) and rotations (up to 30$^\circ$). After applying the transformation, a 2D Fourier transform is performed to obtain the corresponding k-space line. All lines are sequentially stored and combined to reconstruct a motion-corrupted image via inverse FFT. For undersampling, we adopt 1D variable-density Gaussian masks along the phase-encoding direction to mimic clinical acceleration settings. Two acceleration factors are evaluated: $R=4\times$ and $R=8\times$. The Gaussian masks are centered and smoothly decay toward high-frequency regions to preserve low-frequency information. The same sampling pattern is applied consistently across all slices.

\begin{figure}[t]
\centerline{\includegraphics[width=\columnwidth]{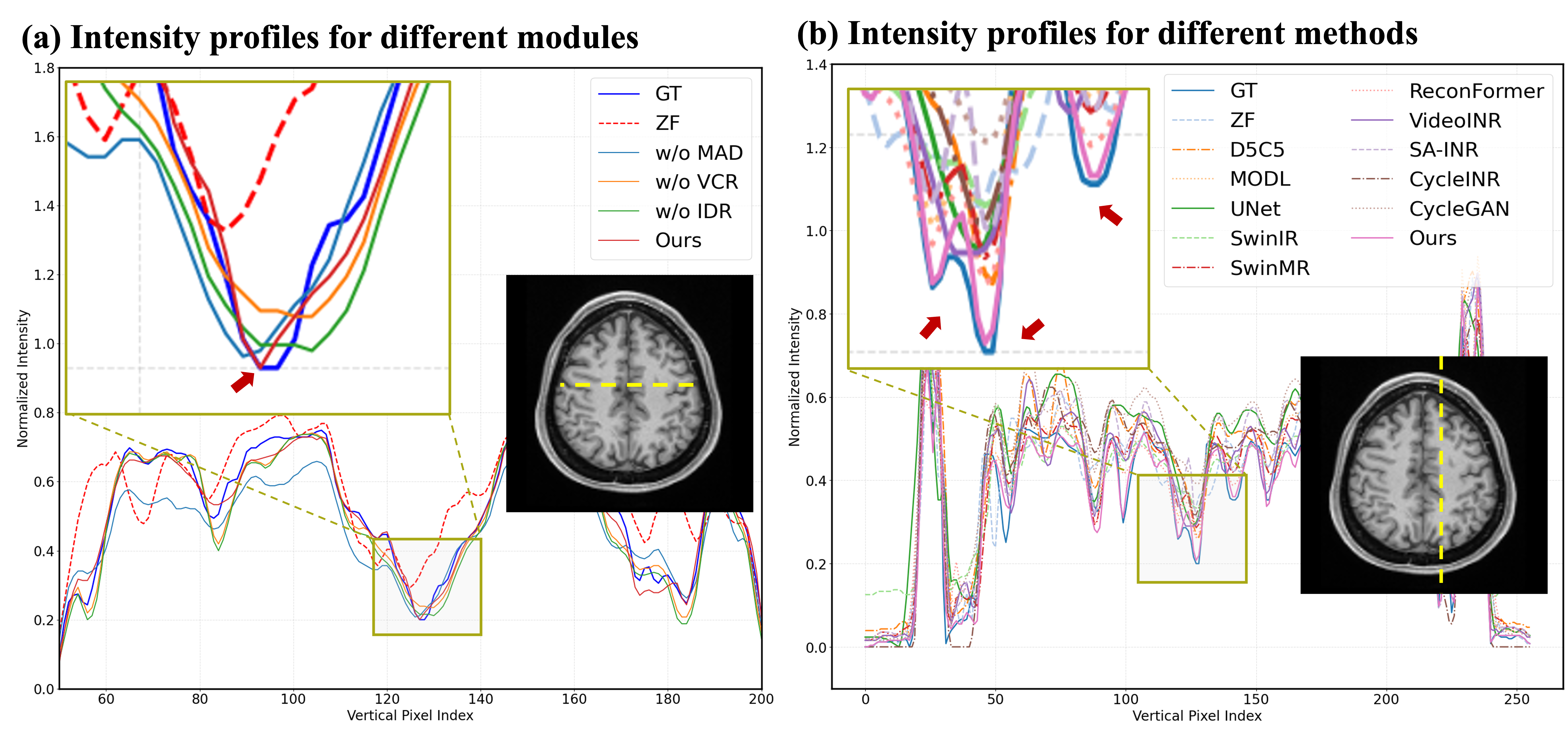}}
\caption{Vertical intensity profile comparison across different reconstruction methods. The plot compares the normalized intensity values of the ground truth (GT), zero-filled (ZF), and various reconstruction approaches. The magnified regions highlight differences in edge fidelity and fine-structure.}
\label{fig7}
\end{figure}

\begin{figure}[t]
\centerline{\includegraphics[width=\columnwidth]{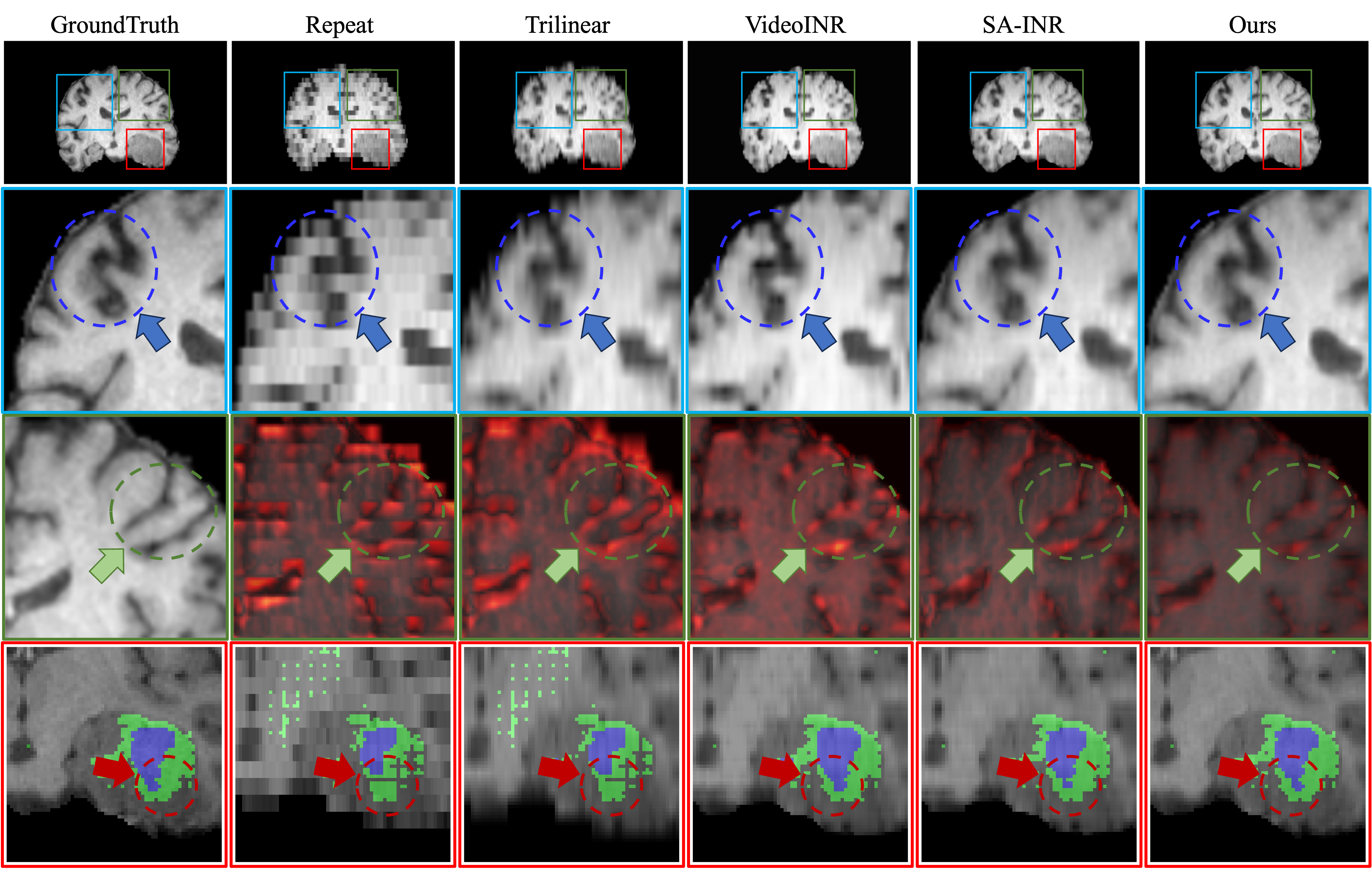}}
\caption{{Visualization of different 3D reconstruction methods. The first row shows the coronal view with selected regions. The second row highlights image reconstruction quality of a coronal view from a stack of thick axial slices. The third row shows the error maps of each methods compared to GroundTruth. The fourth row presents corresponding BraTS tumor segmentation results.}}
\label{fig8}
\end{figure}

\subsection{Metric and Network Configuration}
First, we evaluate PR-INR using several key metrics. For image reconstruction, it employs: 1) PSNR to measure the fidelity of reconstructed images, with higher values indicating less corruption; 2) SSIM to quantify perceptual similarity by analyzing structural, luminance, and contrast features; 3) LPIPS is a learned perceptual metric \cite{lpips} to assess similarity in feature space, capturing fine-grained visual details; 4) NCC to evaluate the correlation between the reconstructed image and the ground truth (GT), with higher values indicating stronger agreement. For the downstream image segmentation, we used the pre-trained UNet network as the baseline \cite{unet}. It employs the Dice score coefficient (DSC) to evaluate overlap between predicted and reference masks. 

Second, the comparison of our PR-INR with eight state-of-the-art MRI reconstruction methods includes visual analysis of multi-datasets, multi-slice analysis, and quantitative analysis. Among them, seven are the fast MRI reconstruction methods \cite{D5C5,MoDL,swinir,unet,swinmr,reconformer,Cyclegan} and three are the INR-based multi-slice reconstruction methods \cite{videoinr,sainr,cycleinr}. In addition, we also included comparisons of some traditional methods, such as zero-filling (for under-sampling), cubic interpolation, and trilinear interpolation (for slice interpolation).

The configuration of our experiment is as follows: {The adaptive moment (Adam) method is applied to minimize the combined error of our PR-INR. The initial learning rate is 0.002, decreasing by a factor of 0.5 every 10 epochs. Training is performed for a maximum of 100 epochs, with early-stopping (patience = 5) monitoring the validation PSNR and dropout (dropout rate = 0.2) to reduce model overfitting.  The batch size is set to 2 due to the high memory requirements of 3D MRI data and the coordinate-based INR representation, which involves full-volume evaluations during training. A weight decay of \(1\times10^{-3}\) is applied for regularization, and gradient clipping with a maximum norm of 1.0 is used to stabilize optimization. All experiments are implemented in PyTorch 2.1 with mixed-precision training (AMP) to improve computational efficiency. The total parameter count of PR-INR is 28.5M.} All calculations were performed on a workstation with a Xeon 2.90 GHz CPU and an NVIDIA RTX A6000 GPU.

\begin{figure}[t]
\centerline{\includegraphics[width=\columnwidth]{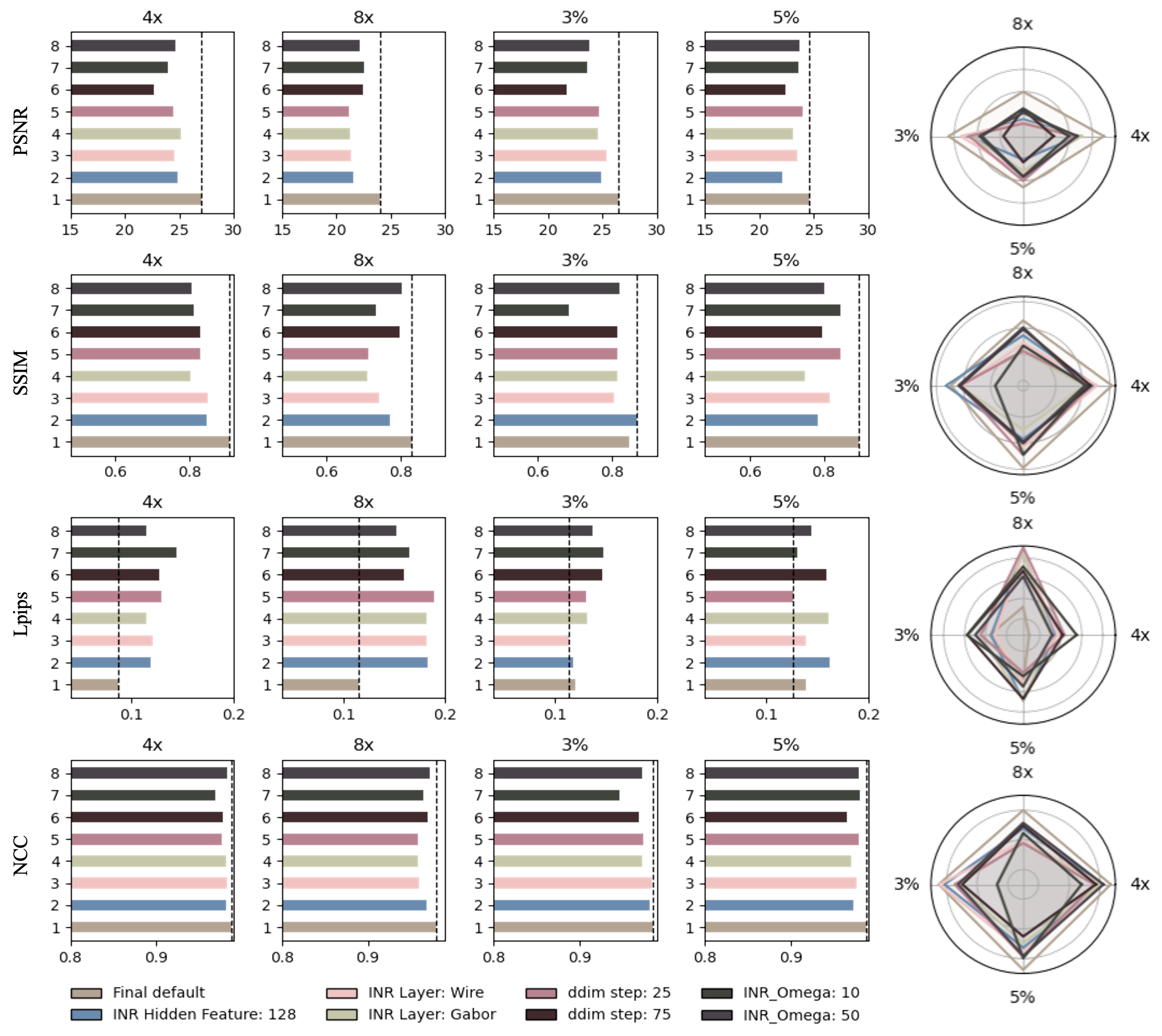}}
\caption{This figure presents the results of an ablation study evaluating different network settings. Radar and bar plots show performance across different settings for INR layers, hidden dimensions, $\Omega$ values in INR network, and DDIM steps.}
\label{fig9}
\end{figure}
\begin{figure*}[t]
\centerline{\includegraphics[width=2\columnwidth]{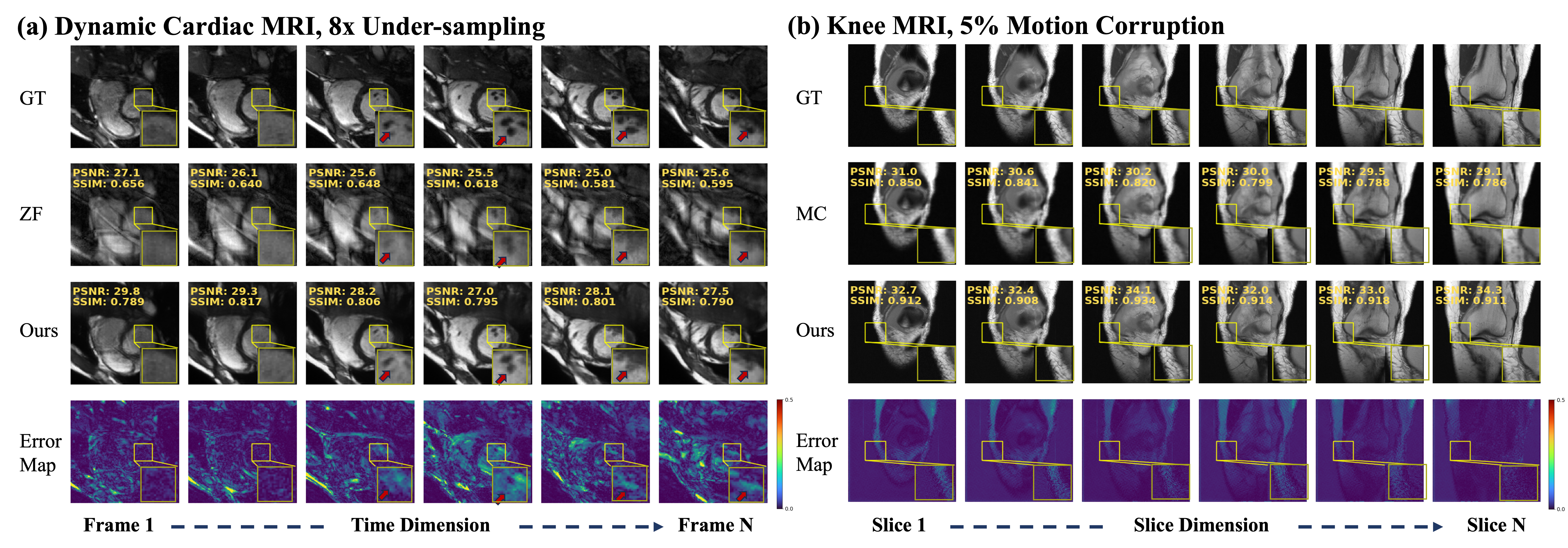}}
\caption{Visual results on unseen domains. (a) cardiac cine MRI across the temporal dimension. 'ZF' represents zero-filled image. (b) knee MRI across the slice dimension. 'MC' represents motion-corrupted image. PR-INR demonstrates improved recovery of anatomical structures and better suppression of undersampling and motion artifacts.}
\label{fig10}
\end{figure*}

\begin{figure}[t]
\centerline{\includegraphics[width=\columnwidth]{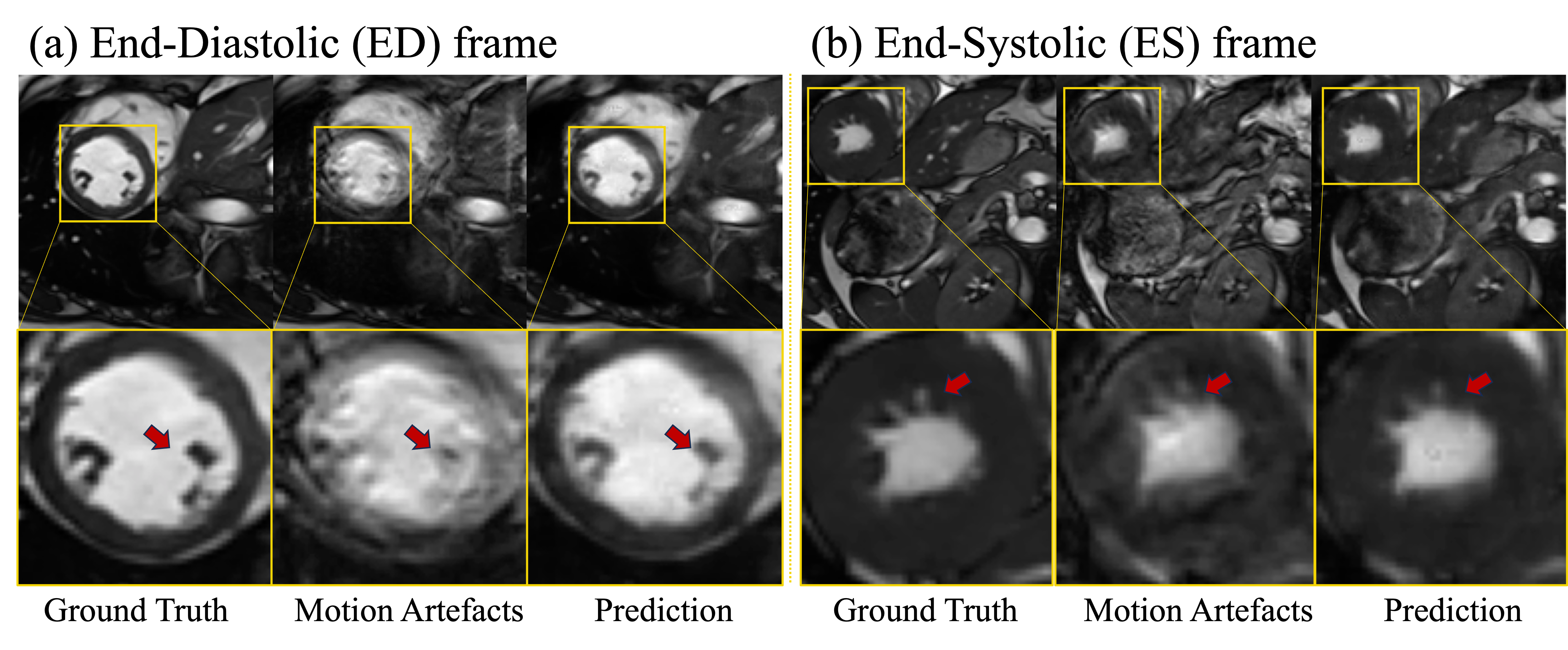}}
\caption{\textcolor[RGB]{0,0,0}{Predicted cardiac MR images at end-diastolic (ED) and end-systolic (ES) frames. Each column shows the ground truth, images with respiration-induced motion artifacts, and the predictions from the proposed PR-INR model, respectively, illustrating the structural differences and motion correction across cardiac phases.}}
\label{fig11}
\end{figure}
\begin{table}[t]
\centering
{
\caption{Quantitative comparison of computational complexity and inference efficiency. The proposed PR-INR is decomposed into its three stages (MAD, IDR, VCR) for detailed analysis.}
\resizebox{\columnwidth}{!}{
\begin{tabular}{l|cccc}
\hline
\textbf{Method} & \textbf{Params (M)} & \textbf{FLOPs (G)} & \textbf{Time (s/slice)} & \textbf{Time (s/volume)} \\
\hline
SwinMR \cite{swinmr} & 11.402 & 800.733 & 0.1884 & 24.11 \\
ReconFormer \cite{reconformer} & 2.822 & 695.796 & 0.0470 & 6.02 \\
VideoINR \cite{videoinr} & 8.101 & 584.283 & 0.2385 & 30.53 \\
SA-INR \cite{sainr} & 2.110 & 59.294 & 0.2295 & 29.38 \\
FCB \cite{FCB} & 12.103 & 26.852 & 0.0642 & 8.25 \\
MMR-Mamba \cite{MMR-Mamba} & 6.805 & 75.140 & 0.0819 & 10.40 \\
\hline
PR-INR (MAD) & 6.850 & 211.098 & 0.4479 & 57.33 \\
PR-INR (MAD+IDR) & 7.131 & 215.511 & 0.4500 & 57.60  \\
PR-INR (MAD+IDR+VCR) & 7.471 & 226.592 & 0.4591 & 58.76 \\
\hline
\end{tabular}}
\label{tab:complexity}}
\end{table}

\subsection{Comparison Experiment}
%{}
\subsubsection{Comparison Experiment in Motion Correlation}
{Table~\ref{tab:motion} demonstrates that our PR-INR effectively compensates for patient motion during k-space acquisition. Compared with existing approaches, our method consistently achieves superior reconstruction quality under different motion corruption setting (3$\%$ and 5$\%$).} It delivers the best quantitative performance, attaining a PSNR of 33.3\,dB and an SSIM of 0.919 when the displacement magnitude is 3\,\% of the FOV and still preserving a 31.0\,dB / 0.887 at the more challenging 5\,\% of the FOV. These results indicate pronounced suppression of aliasing artifacts and a notable enhancement in volumetric spatial coherence. Visualization results (Fig.~\ref{fig4} (a) and (b)) demonstrate that PR-INR restores subtle cortical folds and yields sharper tissue interfaces. The zoomed-in regions further highlight its superiority in reconstructing fine anatomical details and preserving global consistency.{
Besides, a slight upward shift in global intensity can be seen relative to the ground truth. This effect may arises from the residual refinement behavior of the INR module, which predicts high-frequency corrections on top of diffusion- and data-consistent reconstructions. In regions with weak or uncertain structural evidence, the residual estimation may slightly overcompensate the signal magnitude, resulting in a mild global brightness increase after normalization. Moreover, the diffusion prior inherently enhances mid- and high-frequency contrast, which further contributes to this effect. This reflects the model’s intrinsic tendency to emphasize fine-detail recovery and contrast enhancement without compromising anatomical fidelity.}
{Table~\ref{tab:brats} shows that our PR-INR consistently outperforms state-of-the-art MRI reconstruction methods under both 3$\%$ and 5$\%$ motion corruption on the BraTS dataset. It attains the best overall performance with PSNR = 32.1 dB / 30.0 dB, SSIM = 0.962 / 0.942, and NCC = 0.991 / 0.985 for 3$\%$ and 5$\%$ corruption levels, respectively. These results indicate that PR-INR effectively preserves fine structural boundaries and minimizing perceptual distortion (lowest LPIPS = 0.046 / 0.067) from motion-corrupted, lesion-containing MR images.}

\subsubsection{Comparison Experiment in Under-sampled MRI Reconstruction}
{Table~\ref{tab:motion} shows that our PR-INR outperforms state-of-the-art MRI reconstruction methods under both 4× and 8× acceleration on the fastMRI and HCP1200 datasets.} The model maintains superior image quality and structural consistency despite reduced sampling, indicating its capacity to restore high-frequency details and suppress undersampling artifacts. It achieves the highest PSNR (28.9/26.4 dB on fastMRI, 26.4/24.0 dB on HCP1200) and SSIM (0.887/0.721 on fastMRI, 0.895/0.849 on HCP1200), indicating superior reconstruction quality and perceptual fidelity. Fig.~\ref{fig4} (c) and (d) further demonstrate the advantages of PR-INR in visual quality. Compared to baseline models, our method reconstructs sharper anatomical boundaries and suppresses noise and artifacts. The zoomed-in regions especially highlight its ability to retain fine-grained features and maintain consistency with ground truth images.
{Table~\ref{tab:brats} contains the comparison under 4x and 8x undersampling conditions on the BraTS dataset. Our PR-INR demonstrates consistent superiority across all metrics. Specifically, PR-INR achieves PSNR = 34.1 / 29.5 dB, SSIM = 0.984 / 0.935, and LPIPS = 0.020 / 0.045 for 4x and 8x accelerations, respectively. These improvements confirm that PR-INR can maintain structural coherence and tumor visibility even under aggressive undersampling. The high NCC values (0.996 / 0.984) further demonstrate the model’s ability to preserve intensity correlation and fine structural details.}

\subsubsection{Comparison Experiment in Slice Interpolation}
Table~\ref{compar_3D} reports the performance of different inter-slice reconstruction methods and downstream brain tumor segmentation (scale = 5). Our PR-INR achieves the highest PSNR (34.4) and SSIM (0.941), indicating superior image fidelity compared to interpolation baselines. In downstream segmentation, PR-INR also surpasses other methods in capturing anatomical boundaries (69.4  for ET, 87.0 for TC, and 82.6 for WT). Fig. \ref{fig8} further illustrates the differences among reconstruction methods.  Notably, the blue dashed regions reveal that PR-INR maintains structural coherence in low-resolution slices. {The green dashed regions show error maps, where PR-INR exhibits lower reconstruction errors and better preservation of cortical folding patterns, especially along sulcal and gyral boundaries.} The red dashed segmented regions show that PR-INR is spatially consistent and aligned with the ground truth in tumor boundaries (red circles area). These results demonstrate that PR-INR not only improves visual quality but also preserves detailed structure.
\begin{table}[t]
\centering
\caption{Generalization performance on unseen domains.}
\resizebox{\columnwidth}{!}{
\begin{tabular}{c|c|cccc}
\hline
\textbf{Region} & \textbf{Method} & PSNR$\uparrow$ & SSIM$\uparrow$ & LPIPS$\downarrow$ & NCC$\uparrow$ \\
\hline
\multirow{3}{*}{\textbf{Cardiac}} 
& ZF & 25.0 (2.5) & 0.618 (0.061) & 0.340 (0.045) & 0.889 (0.048) \\
& PR-INR & 28.1 (1.9)  & 0.763 (0.051) & 0.217 (0.045)  & 0.948 (0.031)  \\
& Gain & +3.1 & +0.145 & - 0.123 & +0.059 \\
\hline
\multirow{3}{*}{\textbf{Knee}} 
& ZF &28.7 (3.1)  &0.720 (0.101)  & 0.355 (0.068)  & 0.963 (0.030)  \\
& PR-INR &35.6 (2.7)  & 0.906 (0.036) & 0.026 (0.010)  & 0.992 (0.007)  \\
& Gain & +6.9  & +0.186 & –0.329  & +0.029 \\
\hline
\end{tabular}
}
\label{generalization}
\end{table}

\begin{table}[t]
\centering
\color[RGB]{0,0,0}{
\caption{Reconstruction performance on motion-corrupted cardiac MR images at end-diastolic (ED) and end-systolic (ES) frames.}
\resizebox{\columnwidth}{!}{
\begin{tabular}{c|c|cccc}
\hline
\textbf{Cardiac Phase} & \textbf{Method} & PSNR$\uparrow$ & SSIM$\uparrow$ & LPIPS$\downarrow$ & NCC$\uparrow$ \\
\hline
\multirow{3}{*}{\textbf{ED Frame}} 
& Motion-corrupted & 19.9 (3.3) & 0.540 (0.146) & 0.353 (0.081) & 0.750 (0.154) \\
& PR-INR (Ours) & 24.0 (3.1) & 0.726 (0.088) & 0.285 (0.054) & 0.903 (0.059) \\
& Gain & +4.1 & +0.186 & –0.068 & +0.153 \\
\hline
\multirow{3}{*}{\textbf{ES Frame}} 
& Motion-corrupted & 20.0 (3.4) & 0.548 (0.064) & 0.349 (0.048) & 0.885 (0.050) \\
& PR-INR (Ours) & 23.8 (2.7) & 0.727 (0.078) & 0.285 (0.051) & 0.904 (0.055) \\
& Gain & +3.8 & +0.179 & –0.064 & +0.019 \\
\hline
\end{tabular}
\label{VIII}}}
\end{table}

\subsection{Effectiveness Evaluation}
\subsubsection{Visualization Analysis of Different Modules}
Fig.~\ref{fig5} illustrates the working mechanism of the proposed PR-INR framework across its three stages. {In Fig.~\ref{fig5}(a), the motion-aware diffusion module progressively removes motion artifacts through iterative denoising, transforming distorted inputs into anatomically consistent images. Fig.~\ref{fig5}(b) shows the implicit detail restoration (IDR) process, where the conditioned INR enhances local contrast and recovers fine structural details missing in the zero-filled reconstruction. Fig.~\ref{fig5}(c) illustrates the voxel continuous-aware representation (VCR), which refines inter-slice transitions and enforces volumetric continuity, mitigating anisotropy and preserving coherent 3D anatomy.} Fig.~\ref{fig6} illustrates the hybrid reconstruction strategy in the frequency domain, where optimal performance is achieved by allocating the high-frequency region to INR and the low-frequency region to diffusion. {As shown in the left panel, varying the ratio parameter $\rho$ adjusts the relative contribution of each model, resulting in a smooth trade-off between global structure restoration and fine-detail recovery.  The right panel visualizes the corresponding error maps, revealing a clear transition from smoother but oversmoothed reconstructions at low $\rho$ to sharper but noisier outputs at high $\rho$. This observation confirms that diffusion and INR play complementary roles: diffusion excels in denoising and large-scale consistency, while INR contributes to high-frequency refinement and textural fidelity. A balanced frequency allocation between the two yields the most faithful and perceptually coherent reconstruction.}
Fig.~\ref{fig7} shows that our PR-INR achieves the closest alignment with ground-truth intensity profiles{, preserving sharper boundaries and more consistent intensity transitions.  The ablation results in Fig.~\ref{fig7}(a) demonstrate that removing either MAD or IDR degrades boundary sharpness, while the full model maintains stable intensity gradients.  In Fig.~\ref{fig7}(b), PR-INR outperforms recent diffusion- and transformer-based methods by achieving both edge fidelity and smooth intensity continuity.}
\subsubsection{Numerical Evaluation of Different Modules}
Table~\ref{tab:ablation_final} shows that the full configuration of our PR-INR framework consistently achieves the best performance across all undersampling settings on both the FastMRI and HCP 1200 datasets. At 3$\%$ and 5$\%$ displacement during acquisition, the complete model achieves strong gains over the baseline. For instance, PSNR/SSIM improves from 30.0/0.853 to 33.3/0.919 on FastMRI, and from 24.3/0.820 to 27.0/0.904 on HCP 1200 at 3$\%$, with similar trends at 5$\%$. Under 4x and 8x undersampling, PR-INR remains robust, consistently outperforming variants. At 8x, it improves PSNR from 23.9 to 26.4 and SSIM from 0.645 to 0.721 on FastMRI, and shows comparable improvements on HCP. The IDR module plays a key role in enhancing perceptual and structural quality (e.g., LPIPS drops and SSIM improves), while the VCR module further enforces volumetric smoothness, reflected in higher NCC values. These results validate the complementary contributions of each module and the generalizability of PR-INR across different datasets and acceleration regimes.

\subsubsection{Effectiveness Evaluation of Network Configuration}
\textcolor[RGB]{0,0,0}{As shown in Table~\ref{tab:loss_metrics}, the three loss terms contribute in complementary ways. Under 5$\%$ sampling, reducing the MAD weight $\lambda_1$ from 0.5 to 0.3 leads to a drop in PSNR from 31.3,dB to 30.7,dB and a decrease in SSIM from 0.868 to 0.855, indicating insufficient motion suppression. Increasing the IDR weight $\lambda_2$ to 2.0 slightly improves perceptual quality (LPIPS 0.133 vs.\ 0.136) but also results in lower PSNR and SSIM, reflecting a trade-off between detail enhancement and global fidelity. Similarly, reducing the VCR weight $\lambda_3$ from 0.3 to 0.1 causes noticeable degradation in volumetric consistency (NCC decrease from 0.984 to 0.975). Joint perturbations of multiple loss weights further amplify these effects.}
The final default configuration of our PR-INR achieves superior performance across tasks and evaluation metrics (Fig.  \ref{fig9}). For instance, in the 4x setting, it reaches a PSNR of 27.0, representing a gain of 2.5 dB over the 24.5 (layer: wire). In the more challenging 8× setting, the final default reaches a PSNR of 24.00, which is 2.69 dB higher than the next best. It also achieves the best LPIPS (0.115) and SSIM (0.829) among all configurations. Under the motion corruption setting  (3$\%$ and 5$\%$), the final default setup shows clear robustness, reaching up to 26.48 PSNR and 0.847 SSIM at 3$\%$ displacement. These results validate that the default architecture improves perceptual fidelity and structural consistency.

{
\subsubsection{Computational Efficiency}
Table~\ref{tab:complexity} presents the computational complexity and inference efficiency of different reconstruction models. The proposed PR-INR requires 7.47M parameters and 226.6G FLOPs in its full version, which is lighter than SwinMR (11.4M, 800.7G) and comparable to MMR-Mamba (6.8M, 75.1G). The inference time is 0.459\,s per slice, slightly slower than transformer-based models such as SwinMR (0.188\,s/slice) and ReconFormer (0.047\,s/slice). This moderate increase in runtime is acceptable given the substantial improvement in reconstruction fidelity and the inherently slow MRI acquisition process \cite{guenette2024time}.}

\subsection{Generalization Evaluation}
Fig. \ref{fig10} and Table~\ref{generalization} show that the PR-INR framework can be applied to two unseen domains. It includes cardiac-cine (balanced Steady State Free Precession, bSSFP) with multi-frame sequences and knee proton density (PD). Fig. \ref{fig10}(a) shows that PR-INR can preserve ventricular and papillary structures with sharp, temporally consistent boundaries. The error maps  highlight the effectiveness in suppressing aliasing, particularly around the basal septum (red arrows). Fig. \ref{fig10}(b) shows that PR-INR restores cartilage interfaces and meniscal detail lost in motion-corrupted images. Table~\ref{generalization} reports quantitative gains: PR-INR improves PSNR by +3.1/6.9 dB, SSIM by +0.145/0.186, and NCC by +0.059/0.029, while reducing LPIPS by 0.123/0.329 on the cardiac/knee datasets. It demonstrates robust generalization across anatomies and contrasts.

\textcolor[RGB]{0,0,0}{For real motion corruption, we further validated PR-INR on the CMRxMotion dataset containing real respiration-induced motion artifacts \cite{wang2022extreme}. Fig.~\ref{fig11} shows motion-corrupted inputs at end-diastolic (ED) and end-systolic (ES) frames contain real respiration-induced motion artifacts, leading to visible blurring and ghosting around the ventricular walls. The proposed PR-INR model effectively restores structural integrity and sharpness, yielding clearer myocardial boundaries and consistent ventricular morphology across cardiac phases. Table~\ref{VIII} further reports consistent improvements in PSNR (+4.1 and +3.8 dB), SSIM (+0.186 and +0.179) over motion-corrupted inputs for ED and ES frames, confirming the model’s robustness in correcting real respiration-induced motion artifacts.}

\section{Discussion}
The design of PR-INR is explicitly guided by three core hypotheses. It includes motion corruption, hallucination correction, inter-slice continuity and domain generalization. Experiment results provide empirical support for these hypotheses and demonstrate their complementary roles in producing reliable reconstructions.

The first hypothesis posits that a diffusion-based prior can effectively suppress global aliasing and motion-induced artifacts by leveraging its generative capacity. This is supported by our quantitative results under synthetic motion settings (Table~\ref{tab:motion}), where PR-INR achieves 33.3 dB PSNR and 0.919 SSIM at 3\% displacement, and maintains 31.0 dB / 0.887 even under 5\% displacement. Qualitative visualizations (Fig.~\ref{fig4}(a) and (b)) further show that our MAD module restores global cortical folds and large-scale tissue integrity. Fig.~\ref{fig5}(a) illustrates this denoising process over diffusion steps, clearly showing the recovery of anatomical structure. These results validate that the MAD module captures low-frequency structure and motion patterns. However, residual blur and hallucinated textures are still observed around high-frequency regions.

To address this, our second hypothesis posits that hallucinations are spatially and frequency localized and can therefore be selectively refined. This is supported by the error map analysis in Fig.~\ref{fig6}, which shows substantially reduced errors after refinement. Further evidence is provided by the intensity profiles in Fig.~\ref{fig7}, where our method closely aligns with ground truth at structural boundaries, while competing methods exhibit oversmoothing or spurious oscillations. These results indicate that the IDR module effectively suppresses high-frequency hallucination-prone regions by conditioning on coarse outputs and acquisition-aware features.
While this dual-domain refinement mitigates localized hallucination artifacts, residual uncertainty may still arise under severely degraded inputs, such as low-SNR or motion-corrupted slices. To constrain generative behavior and preserve anatomical fidelity, PR-INR integrates dual-domain supervision, volumetric regularization, and strict data-consistency projection. Future work will explore uncertainty quantification and radiologist reader studies to further assess hallucination risk and clinical reliability.

The third hypothesis focuses on the need to enforce volumetric continuity in the presence of anisotropic slice resolution. Our volumetric continuous representation (VCR) module addresses this by modeling a continuous function over 3D coordinates. Experimental results in Table~\ref{compar_3D} and Fig.~\ref{fig8} show that PR-INR achieves the best slice interpolation (PSNR = 34.4, SSIM = 0.941) and yields superior segmentation accuracy (ET: 69.4, TC: 87.0, WT: 82.6). Fig.~\ref{fig5}(c) further illustrates the volumetric modeling process, where spatial continuity is preserved across slices.

Together, these three hypotheses form the foundation of a robust framework. This design not only promotes task-specific optimization but also facilitates generalization across domains. Fig.~\ref{fig10} and Table~\ref{generalization} show that PR-INR achieves up to +6.9 dB PSNR and –0.329 LPIPS improvement on unseen cardiac and knee datasets.

\textcolor[RGB]{0,0,0}{
However, the proposed formulation targets anisotropic multi-slice 2D MRI acquisitions with thick slices. In this setting, motion corruption is dominated by in-plane components. Motion correction in the MAD stage is therefore performed in a slice-wise manner, and acquisition-level slice coupling induced by through-plane encoding and slice profile effects is not explicitly modeled. This provides a practical approximation under thick-slice sampling but does not fully capture inter-slice coupling in more isotropic or fully encoded 3D acquisitions. Future work may incorporate more acquisition-consistent volumetric motion models (e.g., joint optimization across slices) to better model inter-slice coupling within a continuous implicit representation framework.
}

\section{Conclusion}
In conclusion, we propose a progressive refinement implicit neural representation (PR-INR) framework for MRI slice-to-volume reconstruction. The framework combines motion-aware diffusion, INR-based detail restoration, and continuous volumetric modeling to address motion artifacts, recover high-frequency details, and enforce slice-to-slice consistency under anisotropic sampling. Extensive experiments on four public datasets demonstrate consistent improvements in both quantitative metrics and visual quality. We further validate the versatility of PR-INR on cardiac and knee imaging data, highlighting its robustness and potential for broader adoption in medical imaging workflows.

\section{Appendix}
\label{appendix}
\subsection{Proof and Analysis for Equation 14}

The MRI reconstruction pipeline consists of two stages:
\begin{enumerate}
    \item Diffusion-based Motion Correction: Generates initial estimate $\hat{x}_0$ with motion artifacts mitigated
    \item Data-Consistent INR Refinement: Produces $\xidr$ via the sum up residual predicted by INR:
    \begin{equation}
        \xidr = \xtilde + f_\phi(\xtilde, y, z).
    \end{equation}
\end{enumerate}

Variance Analysis of Stage 2:
\label{sec:var_analysis}

\begin{definition}[k-Space Decomposition]
\label{def:kspace}
Given sampling mask $M \in \{0,1\}^{n\times n}$, we decompose the reconstruction:
\begin{align}
    \xlow &= \F^{-1}(M \odot y),  \\
    \xhigh &= \F^{-1}((1-M) \odot \F(\hat{x}_0)), \quad 
\end{align}
where $\odot$ denotes the Hadamard product.
\end{definition}

After data consistency projection, we have two properties:

Low-frequency determinism: $\Var(\xlow|y) = 0$

High-frequency inheritance: $\Var(x^{\mathrm{high}}|y) = \Var(\hat{x}_0^{\mathrm{high}}|y)$

% For property 1:
% \begin{equation}
%     \xlow = \F^{-1}(M \odot y) \Rightarrow \xlow \text{ is } \sigma(y)\text{-measurable} \Rightarrow \Var(\xlow|y) = 0
% \end{equation}

% For property 2:
% \begin{equation}
%     \xhigh = \F^{-1} \circ (1-M) \odot \F \circ \hat{x}_0 \Rightarrow \text{Linear transformation preserves variance scaling}
% \end{equation}

\subsection{Residual Learning Formalism}
\label{sec:residual}

\begin{definition}[PR-INR Operator]
\label{def:idr}
The refinement operator of IDR process acts as:
\begin{equation}
    \xidr = \xtilde + r, \quad r = f_\phi(\xtilde, y, z).
\end{equation}
with $r$ constrained to the high-frequency subspace:
\begin{equation}
    (\F(r)) \subseteq \{ (k_x,k_z) | M(k_x,k_z)=0 \}
\end{equation}
\end{definition}

\begin{theorem}[Variance Decomposition]
\label{thm:var_decomp}
The conditional variance decomposes as:
\begin{equation}
    \Var(\xidr|y) = \underbrace{\Var(\xhigh|y)}_{T_1} + \underbrace{\Var(r|y)}_{T_2} + \underbrace{2\Cov(\xhigh, r|y)}_{T_3}.
\end{equation}
\end{theorem}

$\text{(since } \Var(\xlow|y)=0 \text{)}$

\subsection{Variance Reduction Mechanism in IDR}
\label{sec:mechanism}

\begin{lemma}[Dual-Domain Constraint]
\label{lem:dual_loss}
The composite loss:
\begin{equation}
    \mathcal{L}_{\mathrm{IDR}} = \lambda_1\|\xidr - x_{\mathrm{gt}}\|^2 + \lambda_2\|P \odot \F(\xidr) - y\|^2.
\end{equation}

induces the variance bound:

\begin{equation}
    \Var(r|y) \leq \frac{1}{\lambda_1 + \lambda_2}\E\left[\left({\mathcal{L}}{r}\right)^2\right]
\end{equation}
\end{lemma}

Via Fisher information inequality:
\begin{equation}
    \Var(r) \leq \frac{\E[(\partial_r \mathcal{L})^2]}{(\partial^2_r\mathcal{L})^2} \leq \frac{\E[(\partial_r \mathcal{L})^2]}{(\lambda_1 + \lambda_2)^2}.
\end{equation}
where second derivative $\partial^2_r \mathcal{L} \geq \lambda_1 + \lambda_2$.

\begin{theorem}[Dominance of Covariance Term]
\label{thm:cov_dom}
Under optimal training:
\begin{equation}
    \Cov(\xhigh, r|y) \leq \sqrt{\Var(\xhigh|y)\Var(r|y)}.
\end{equation}
\end{theorem}

Using Cauchy-Schwarz inequality:
\begin{equation}
\begin{aligned}
    \Cov(\xhigh, r) &= \E[(\xhigh - \E\xhigh)(r - \E r)] \\
    &\leq -\sqrt{\Var(\xhigh)\Var(r)}.
\end{aligned}
\end{equation}
\text{(by error correction property)}. The negative sign arises because $r$ actively corrects errors in $\xhigh$.

\subsection{Main Variance Reduction Theorem}
\label{sec:main_result}

\begin{theorem}[Strict Variance Reduction]
\label{thm:main}
Under the pipeline constraints:
\begin{equation}
    \Var(\xidr|y) \ll \Var(\hat{x}_0|y).
\end{equation}
\end{theorem}

Combine results from \cref{thm:var_decomp,lem:dual_loss,thm:cov_dom}:
\begin{equation}
\begin{aligned}
\Var(\xidr|y) &= T_1 + T_2 + T_3 \\
&= \Var(\xhigh|y) + \Var(r|y) + 2\Cov(\xhigh, r|y) \\
&\leq \Var(\xhigh|y) - 2\sqrt{\Var(\xhigh|y)\epsilon} + \epsilon \\
&= \left( \sqrt{\Var(\xhigh|y)} - \sqrt{\epsilon} \right)^2 \\
&\ll \Var(\hat{x}_0|y).
\end{aligned}
\label{eq:variance_bound}
\end{equation}

where $\epsilon = \Var(r|y)$ is constrained by the dual-domain loss.

\subsection{Proposition (Discrete Smoothness of INR in the Axial Direction).}

Let \( f_\theta : \mathbb{R}^3 \to \mathbb{R} \) be a coordinate-based implicit neural representation (INR), parameterized by \( \theta \), which maps continuous spatial coordinates \( (x, y, z) \in \mathbb{R}^3 \) to scalar intensity values. Assume that:

\begin{itemize}
    \item \( f_\theta \) is implemented as a multilayer perceptron (MLP) with continuous activation functions (e.g., ReLU, tanh, sin),
    \item The network is trained on volumetric data sampled at discrete slice indices \( z_i \in \mathbb{Z} \),
    \item The domain \( \Omega \subset \mathbb{R}^2 \times \mathbb{Z} \) is compact (i.e., the spatial and axial coordinates are bounded).
\end{itemize}

Then, there exists a constant \( C > 0 \) such that for all \( (x, y) \in \mathbb{R}^2 \) and all \( z \in \mathbb{Z} \), the finite forward difference of \( f_\theta \) along the \( z \)-axis is uniformly bounded:

\begin{equation}
\left| f_\theta(x, y, z + 1) - f_\theta(x, y, z) \right| \leq C.
\end{equation}

This inequality expresses a form of discrete Lipschitz continuity over the \( z \)-axis and implies that the learned INR is smooth across adjacent slices in the axial direction.

% \section{References}
\bibliographystyle{IEEEtran}
\bibliography{ref}

\end{document}